\documentclass[traditabstract,usenatbib]{aa}
\usepackage{txfonts}
\usepackage{graphicx}                      
\usepackage{url} 
\usepackage{natbib}               
\bibpunct{(}{)}{;}{a}{}{,}	

\begin{document}
   \title{The Arches cluster out to its tidal radius: dynamical mass segregation and the effect of the extinction law on the stellar mass function. \thanks{Based on observations collected at the ESO/VLT under Program ID 081.D-0572(B) (PI: W.~Brandner) and ID 71.C-0344(A) (PI: F.~Eisenhauer, retrieved from the ESO archive). Also based on data collected at the Subaru Telescope, which is operated by the National Astronomical Observatory of Japan.}}


\author{
M. ~Habibi\inst{\ref{inst1},\ref{inst3}}
, A. ~Stolte\inst{\ref{inst1}}
, W. ~Brandner\inst{\ref{inst2}}
, B. ~Hu{\ss}mann\inst{\ref{inst1}}
, K. ~Motohara \inst{\ref{inst4}}
}
\authorrunning{ Habibi et al.}

\institute{
	   Argelander Institut f\"ur Astronomie, Universit\"at Bonn, Auf dem H\"ugel 71, 53121 Bonn, Germany \email{[mhabibi;astolte;hussmann]@astro.uni-bonn.de}\label{inst1}
	   \and Max-Planck-Institut f\"ur Astronomie, K\"onigsstuhl 17, 69117 Heidelberg, Germany \email{brandner@mpia.de}\label{inst2}
	   \and  Member of the International Max Planck Research School (IMPRS) for Astronomy and Astrophysics at the Universities of Bonn and Cologne. \label{inst3}
	   \and Institute of Astronomy, The University of Tokyo, Osawa 2-21-1, Mitaka, Tokyo 181-0015, Japan \email{kmotohara@ioa.s.u-tokyo.ac.jp}\label{inst4}
	   	   }

  \date{Received Oct 14, 2012; accepted May 13, 2013}

  \abstract
   {The Galactic center is the most active site of star formation in the Milky Way Galaxy, where particularly high-mass stars have formed very recently and are still forming today. However, since we are looking at the Galactic center through the Galactic disk, knowledge of extinction is crucial when studying this region. The Arches cluster is a young, massive starburst cluster near the Galactic center.  We observed  the Arches cluster out to its tidal radius using $K_{s}$-band imaging obtained with NAOS/CONICA at the VLT combined with Subaro/Cisco $J$-band data to gain a full understanding of the cluster mass distribution. We show that the determination of the mass of the most massive star in the Arches cluster, which had been used in previous studies to establish an upper mass limit for the star formation process in the Milky Way, strongly depends on the assumed  slope of the  extinction law. Assuming the two regimes of widely used infrared extinction laws, we show that the difference can reach up to 30\% for individually derived stellar masses and $\Delta A_{Ks}\sim 1$ magnitude in acquired $K_{s}$-band extinction, while the present-day mass function slope changes by $\sim 0.17$ dex. The present-day mass function slope derived assuming the more recent extinction law increases from a flat slope of $\alpha_{Nishi}=-1.50 \pm0.35$ in the core ($r<0.2 $ pc) to $\alpha_{Nishi}=-2.21 \pm0.27$ in the intermediate annulus ($0.2 <r<0.4$ pc), where the Salpeter slope is -2.3. The mass function steepens to $\alpha_{Nishi}=-3.21 \pm0.30$ in the outer annulus ($0.4<r<1.5 $ pc), indicating that the outer cluster region is depleted of high-mass stars. This picture is consistent with mass segregation owing to the dynamical evolution of the cluster.
   
 }


\titlerunning{The Arches cluster out to its tidal radius: dynamical mass segregation and the effect of the extinction law.}
\authorrunning{M. Habibi et al.}

   \maketitle
%

\section{Introduction}

The center of our galaxy hosts many massive molecular clouds, active sites of massive star formation, and a remarkable number of high-mass stars (Yusef-Zadeh et al. \citealt{Yusef-Zadeh2009}). It contains three of the most massive and young star clusters in our Galaxy, namely: the Arches, Quintuplet and the young nuclear cluster. The Arches cluster, located at a projected distance of $\sim 26\;$pc from the Galactic center (GC), is the youngest among the three ($\sim 2.5\;$Myr, Najarro et al. \citealt{Najarro}; Martins et al. \citealt{martins}). This cluster contains many massive stars including  160 O-stars and 13 WN stars (Figer \citealt{figer2004conf} ; Martins et al. \citealt{martins}).  With its collection of high-mass stars and its very dense core ($\rho_{core} \sim 10^{5}$ ${M_{\bigodot}}/{{\rm pc^3}}$; Espinoza et al. \citealt{espinoza}), the Arches cluster is an excellent site for addressing questions about massive star formation, the stellar mass function, and the dynamical evolution of young, massive clusters in the GC environment. 
 
The high extinction toward this  cluster nevertheless causes this investigation to be challenging. A visual extinction of up to 30 magnitudes allows the GC to only be accessed in the radio, infrared, and X-ray regimes. The extinction is caused by the diffuse interstellar medium (Lebofsky \citealt{lybofsky1979}), as well as dense molecular clouds in the central molecular zone (Morris \& Serabyn \citealt{Morris-Serabyn}). The latter are suggested to contribute one-third of the total extinction (Williams et al. \citealt{Williams2000}). The substructure of molecular clouds (e.g. Williams et al. \citealp{Williams2000}) causes patchy extinction along the GC line of sight. The interstellar environment around the Arches cluster itself is also extreme. The Arches cluster is observable in the near infrared thanks to shedding its natal cloud but it is still surrounded by the molecular clouds with which the cluster is interacting (e.g. Yusef-Zadeh et al. \citealp{Yusef-Zadeh2002}; Lang \citealp{lang2004}). 

The line-of-sight extinction toward the GC was first investigated by Becklin \& Neugebauer \cite{Becklin}. The visual to near-infrared extinction law was derived by Rieke \& Lebofsky \cite{rieke1985} by studying the color excess of five supergiants near the GC and the star o Sco in the upper scorpius region. The extinction derived by Rieke \& Lebofsky  \cite{rieke1985} was later fitted to  a power law $A_\lambda \propto \lambda^{-1.61}$ (Cardelli \citealp{cardelli}) and considered in many studies as the standard extinction law toward the GC. Later studies analyzed nebular hydrogen lines (eg. Lutz et al. \citealp{lutz}), near-infrared surveys (Nishiyama et al. \citealp{Nishi2009}; Stead \& Hoare \citealp{Stead}), and the photometry of red clump stars in the GC (Schoedel et al. \citealp{schoedel}). They have revealed  that the near-infrared extinction law in the range of the $J$, $H$, and $K_{s}$ bands is better described by a steeper power law with an index of $\alpha \sim -2.0$.
The slope of the extinction law influences the derivation of individual masses from the luminosities of cluster members, so is expected to influence the shape of the 
stellar mass function and to have especially severe impact on  the  upper mass limit derived from comparison with stellar evolution models.

 Previous studies of the Arches cluster were aware of the high extinction toward this cluster but have chosen different methods of addressing this problem.It is therefore not surprising that different studies resulted in deviating mass functions in the center of the Arches cluster. Figer et al. \cite{figer1999} adopted a single extinction value of $A_{K}=3.0$ mag and found a flat mass function for the Arches cluster compared to the standard Salpeter  \cite{Salpeter} initial mass function. This finding opened a series of debates and further studies with the aim of distinguishing between primordial or dynamical mass segregation.  Stolte et al. \cite{stolte_2002} and \cite{stolte_2005} utilized deep adaptive optics (AO) observations to derive the present-day mass function (PDMF) within the cluster's half-mass radius. They treated  the variable extinction of  the cluster by correcting for a radial reddening gradient  based on the extinction law of Rieke \& Lebofsky \cite{rieke1985}. A study by Kim et al. \cite{kim2006} applied the extinction law of Rieke, Rieke \& Paul \cite{rrp} and chose to correct for a single, mean extinction value for each of the observed cluster annuli and each control field. These studies confirmed the flattening of the mass function in the center of the cluster and also achieved to study the less massive sources in the cluster. Finally, Espinoza et al. \cite{espinoza}  used the general Galactic extinction law by Fitzpatrick \cite{Fitzpatrick} to derive the  mass function. They acquired extinction parameters for the NACO/VLT  natural photometric system and corrected the individual extinction toward each star. As discussed in Sect. \ref{sec:mf}, these studies result in central slopes of $-1.26 < \alpha < -1.88$, and could not come to a common conclusion in regard to the central Arches mass function.

In this paper, we study the effects of the new regime of steeper near-infrared extinction laws on the derivation of individual stellar parameters toward the Arches cluster. The $K_{s}$ and $H$-band AO images taken with VLT/NACO as well as $J$-band images obtained with Subaru/CISCO are analyzed to cover the outskirts  of the Arches cluster for the first time. This paper is organized as follows: in Sect. \ref{sec:obs}, we describe the data and observational setting. The data reduction steps, photometry, and calibration processes are explained in Sect. \ref{sec:DR}. We determine the individual extinction values and construct the extinction map of the cluster in Sect. \ref{sec:exti}. In the last section of Sect. \ref{sec:exti}, the individual stellar parameters derived assuming the two regimes of extinction laws are compared. We investigate the effect of the assumed slope of the extinction law on the determination of the mass of the most massive star in the Arches cluster, which had been used in previous studies to establish an upper mass limit for the star formation process in the Milky Way. In Sect. \ref{sec:mf}, the PDMF of the Arches cluster
\footnote{Some of the previous studies have used the initial masses corresponding to the observed stellar luminosities to construct a presumably initial mass function. The rapid dynamical evolution of massive star clusters such as the Arches (e.g., Kim et al.\citealt{kim2006}, Harfst et al.\citealt{harfst2010}) implies that the observed stellar mass function can be influenced by stellar evolution, as well as by the dynamical evolution of the cluster, so it does not represent the initial mass function even at the current young cluster age. All the mass functions in this study are constructed based on the current spatial distribution of the derived present-day masses derived from stellar evolution models. For the sake of comparison with other studies, we have included the mass function slopes constructed from the initial individual masses, i.e. only corrected for stellar evolution, in Table \ref{slopes}.} is constructed  for an area that reaches closer to the tidal radius of the cluster for the first time. We probe the radial variation in the slope of the mass function in order to distinguish between primordial or dynamical mass segregation in the Arches cluster. A summary of our findings is presented in Sect. \ref{sec:conc}.


\section{Observations}
\label{sec:obs}
\begin{figure}
   \centering
   \includegraphics[scale=0.3]{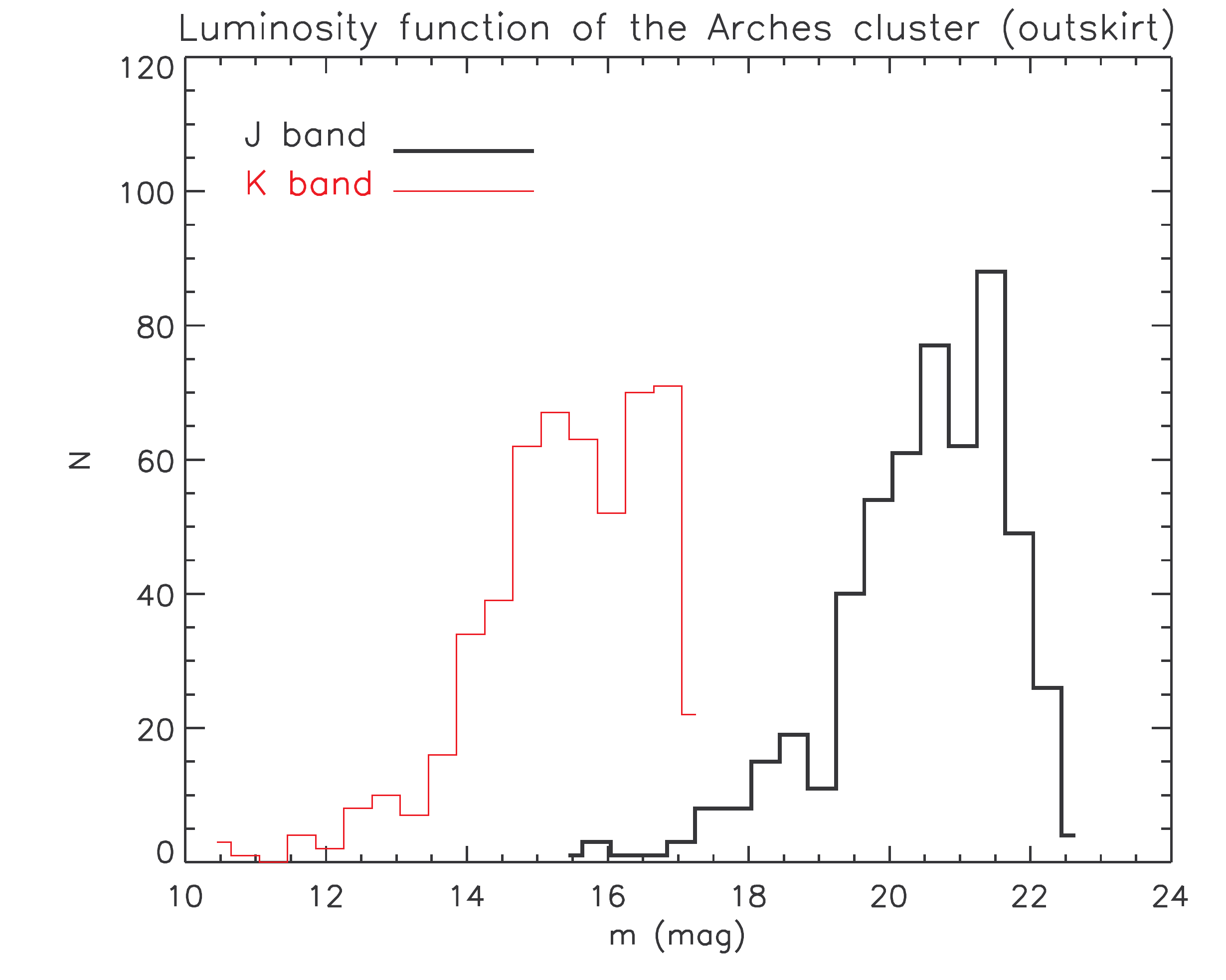}
      \caption{Luminosity functions of all detected sources in the outskirts of the cluster in $J$-band and $K_{s}$-band. The luminosity at which the number of stars decreases represents the detection limit of 21.5 mag and 17 mag on the CISCO  $J$-band and NACO $K_{s}$-band images, respectively. }
	\label{lf}
  \end{figure}

$K_{s}$-band images ($\lambda_{c}=2.18\mu$m, $\triangle\lambda=0.35 \mu$m) of the outer parts of the Arches cluster were obtained on June 6-10, 2008, with the Very
Large Telescope (VLT). Images were acquired with the near-infrared camera CONICA (Lenzen et al. \citealp{lenzen}). The atmospheric turbulence was corrected with the AO system NAOS to enhance the spatial resolution (Rousset et al.  \citealp{Rousset}). We used the brightest star in $K_{s}$ in each field as the natural guide source for the infrared wave-front sensor. Since the available natural guide stars are with $ 9.2<K_{s}<10.4$ mag at the faint limit of the infrared wave-front sensor magnitude range, we had to use the N90C10 dichroic, which distributes 90\% of the light to NAOS, while only 10\% is delivered to the science detector. We used Fowler sampling to enhance the sensitivity that allowed us to detect sources as faint as $K_{s} = 17$ mag (Fig. \ref{lf}). The acquired images cover four fields of $27.8'' \times 27.8''$ each, provided by the medium-resolution camera (S27) with a pixel scale
of 0.027''. On each of the four fields we obtained 44 dithered images with a detector integration time of 15 s and two exposures (NDIT=2), yielding 30 s of total exposure time per frame.
 We refer to the four outer fields as fields two to five. Small dither shifts with a maximum of $2^{''}$ allowed a good background subtraction using the science frames, while at the same time the optical distortions at the edges were minimized. During the $K_{s}$-band observations, the natural visual seeing varied from 0.61 to 0.98 (see Table \ref{ob_log}). We achieved typical spatial resolutions of three to four pixels ($0.081^{''}-0.135^{''}$) on individual frames using this AO setup. The
properties of all fields are summarized in Table \ref{ob_log}.

We used the NAOS-CONICA observations obtained in March 2002 in the central part of the Arches cluster to cover the whole cluster area. A visual guide star with V=16.2 mag was used as the guide probe for the visual wave-front sensor.  The data properties and reduction steps are described in Stolte et al.~\cite{stolte_2005} (see also Stolte \cite{stolte_2003} for details). In this paper, we label this dataset as field 1.

\begin{figure}
   \centering
   \includegraphics[scale=0.3]{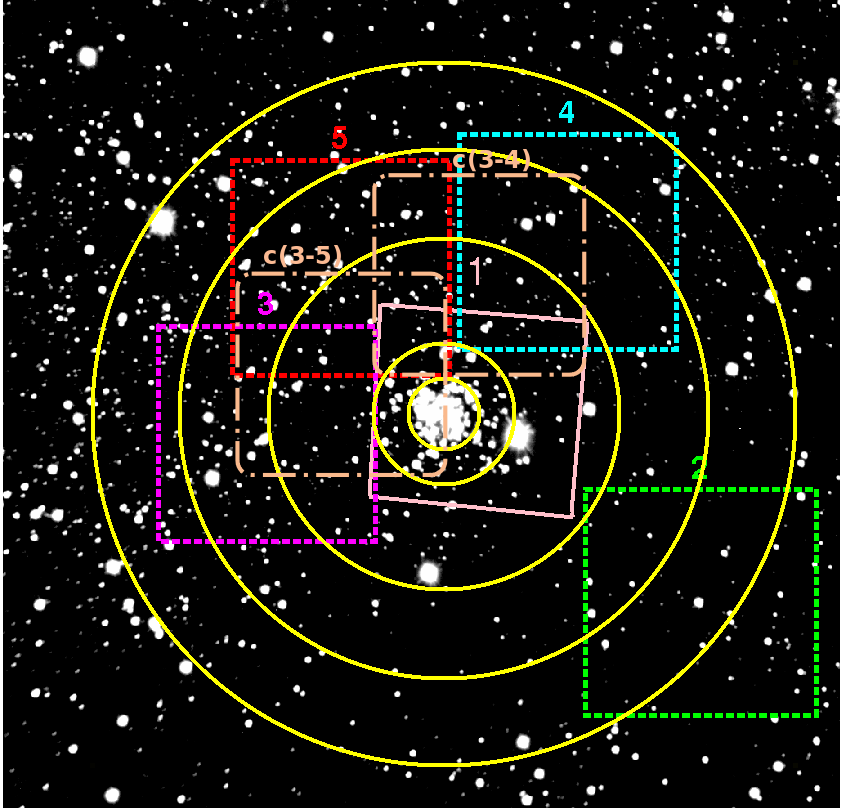}
 \caption{The locations of the NACO fields ($K_{s}$-band observations) overlaid on the Subaru/CISCO $J$-band image of the Arches cluster. Circles illustrate distances of 0.2, 0.4, 1, 1.5, and 2 pc from the center of the cluster. The central field (field 1) was observed in $K_{s}$ and $H$. The two dot-dashed boxes represent the $K_{s}$ calibration frames (see Sect. \ref{sec:calib}). North is up and east to the left.}
\label{fields}
  \end{figure}

Seeing-limited $J$-band ($\lambda_{c}=1.25\mu$m, $\triangle\lambda=0.16 \mu$m) observations were performed on July 17, 2000, with the 8.2
meter Subaru optical-infrared Telescope on Hawai'i (Iye et al.~\citealt{subaru}) under excellent weather conditions. The CISCO spectrograph and camera (Motohara et al.~\citealt{cisco}) provided a pixel scale of 0.116" and a field of view of $2' \times 2'$. We obtained 86 dithered $J$-band images with 10 s exposure time per image.
An average seeing of 0.49" resulted into a full width at half maximum (FWHM) of the point-spread function (PSF) of 0.39'' on the combined image (Table \ref{ob_log}).  The locations of the NACO fields overlaid on the Subaru $J$-band
image are shown in Fig.~\ref{fields}. The luminosity functions (Fig.~\ref{lf}) illustrate that the detection limits on the combined images are $J<21.5$ mag (CISCO) and $K_{s}<17$ mag (NACO).

We also used the Galactic Plane Survey (GPS) of the UKIRT Infrared Deep Sky Survey (UKIDSS) (Lawrence et al. \citealp{ukidds}) and the NACO ground-based photometry of Espinoza et al. \cite{espinoza} for calibration purposes. The UKIDSS survey started in 2005 using the wide field camera on the United Kingdom Infrared Telescope on Mauna Kea, Hawai'i. The GPS covered an area of 1800 square degrees in JHK to a depth of K=19 mag and J=20 mag (Lucas et al. \citealp{Gps}). 

\begin{table*}
\caption{Overview of the observations.}\label{t_observations}
\centering
\begin{tabular}{crrlclccc} 
\hline\hline             
Field & Date & Filter & DIT $\times$ NDIT $\mathrm{(s)}^{\mathrm{(a)}}$  &  $N_{frames}$ & $N_{use}$ & Airmass & Seeing$^{\mathrm{(b)}}$ ($\arcsec$)& FWHM of PSF ($\arcsec$) \\

\hline
  \multicolumn{9}{c}{\it VLT / NACO}\\
\hline
field 1 (center) & 2002-03-30	& $H$ & $30.00\times 2$ 	&	35	& 14	&	1.15 & 0.8 & 0.084 \\
field 1 (center) & 2002-03-30	& $K_{s}$ & $15.00\times 4$ 	&	15	& 7	&	1.15 & 0.8 & 0.084 \\
field 2 (outskirt) & 2008-06-06	& $K_{s}$ & $15.00\times 3$ 	&	44	& 38	&	1.01 & 0.61--0.78 & 0.073 \\
field 3 (outskirt) & 2008-06-06	& $K_{s}$ & $15.00\times 3$ 	&	44	& 32	&	1.06 & 0.69--0.92 & 0.089 \\
field 4 (outskirt) & 2008-06-06	& $K_{s}$ & $15.00\times 3$ 	&	44	& 33	&	1.14 & 0.78--0.98 & 0.079 \\
field 5 (outskirt) & 2008-06-10	& $K_{s}$ & $15.00\times 3$ 	&	44	& 30	&	1.01 & 0.7--0.84 & 0.116 \\

\hline
  \multicolumn{9}{c}{\it Suburu / Cisco} \\
\hline
  wide field & 2000-07-17  & $J$ & $10.00\times 1$ 	&	86 &	72	&	1.72 & -- & 0.39  \\
\hline
\hline
  \multicolumn{9}{c}{\it Calibration  fields} \\
\hline
   calib3-5 & 2012-06-12  & $K_{s}$ & $3.00\times 1$ 	&	88 &	88	&	1.01 & 0.78 & 0.151  \\
\hline
   calib4-5 & 2012-06-12  & $K_{s}$ & $3.00\times 1$ 	&	71 &	71	&	1.72 & 0.9 & 0.108  \\
\hline
\end{tabular}
\tablefoot{
  \tablefoottext{a}{Integration time for each exposure $\times$ number of exposures. }
  \tablefoottext{b}{Optical ($V$-band) seeing. }  
}

\label{ob_log}
\end{table*}

\section{Data reduction}
\label{sec:DR}
We used a self-developed data reduction pipeline to reduce the NACO data. The body of this pipeline was written in Python, and it invokes several IRAF\footnote{IRAF is distributed by the National Optical Astronomy Observatory, which is operated by the Association of Universities for Research in Astronomy (AURA) under cooperative agreement with the National Science Foundation.
} tasks (Tody \citealp{iraf}), as well as custom-made IDL and Python routines. The details of the pipeline data reduction are summarized in Hu{\ss}mann et al. \cite{benjamin}. Images were dark-subtracted with appropriate dark frames for each exposure time and flat fielded using twilight flats. A master dark frame was co-added using three individual dark frames to reduce the noise. To obtain the master flat, a robust linear fit was performed for each pixel on the detector to derive the pixel response at different flux levels. Bad pixels were revealed as pixels whose response after normalizing is less than 0.5 or more than 2. These bad pixels combined with hot and dead pixels detected during the combination of dark frames were written into a bad pixel mask. Ten off-cluster sky frames per field were taken at the end of each one hour observing block. A master sky was created using the IRAF task \texttt{imcombine} from only sky frames in the case of fields 2 and 5, while we acquired the smoothest master sky frame for fields 3 and 4 by combining both science and sky frames. Since it is not possible to find sky regions devoid of stars in the crowded area of the GC, the number of $n_\mathrm{high}=number\: of\: sky\: frames-$ five brightest pixels were rejected to remove the residual star light from the master sky.
 We used the IRAF task \texttt{cosmicray} to reject individual pixels with fluxes eight times above the background standard deviation as cosmic rays in each individual science frame. We added each cosmic-ray mask to the bad pixel mask to obtain individual masks for each science frame.

 Before co-adding the reduced images we calculated the FWHM of a reference source in each field. Combining images with low FWHM provides a higher resolution in the final image while adding a fewer number of images might cause a loss in sensitivity. To avoid degrading  the spatial resolution in the combined images, we selected images with an FWHM of less than $0.0813^{''}-0.1355^{''}$ in dependence on the quality in each field. The number of combined images for each field  are listed in Table \ref{ob_log}. The acquired FWHM values were used to weight the individual images during image combination so that images with low FWHM gained higher weights. 
 
We combined the weighted science images using  the IRAF task \texttt{drizzle} (Fruchter \& Hook  \citealp{Fruchter}). Drizzle applies each individual bad pixel mask to discard hot pixels and other image defects during image combination.

\begin{figure}
   \centering
\includegraphics[scale=0.4]{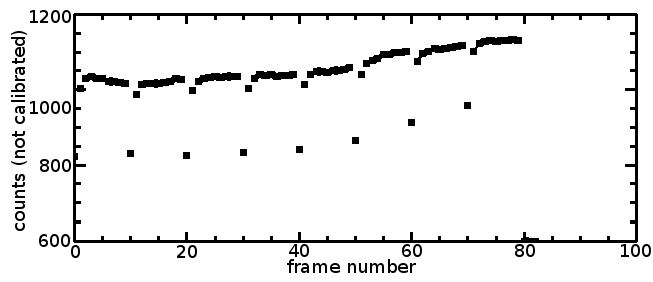}
\caption{The background level of individual frames taken with Subaru/CISCO is plotted in counts. After each shift of the telescope a discontinuity of $\sim 66\%$  was present in the background level of the frames. By scaling images to their background level the trend was removed.  }
\label{suburu-reduce}
\end{figure}

The CISCO data were reduced within IRAF and by employing some custom-made IDL routines. 
Comparing the background level of different frames revealed the obvious trend in discontinuity present in the background (Fig. \ref{suburu-reduce}). The trend was the result of moving and resetting the telescope to capture dithered images so that the background level was down to 66\% in the first three images after each shift \footnote{This instrumental issue is known as ``reset anomaly'' and is well known for many HAWAII arrays (see e.g. http://www.ast.cam.ac.uk/$\sim$optics/cirsi/documents/resetanomaly.html).}. We divided the data according to their background levels into four groups and scaled each group to the mean count of the background level. Scaled images were flat-fielded with twilight flats. Images for sky subtraction and bad pixel masks were created in the same way as for the NACO data. To prepare images for cross-correlation, we applied the IRAF task \texttt{precor} on all science images. Using \texttt{precor}, we removed the majority of cosmic rays and hot pixels by calculating the fraction of pixels in a $3\times3$ pixel box that had a value above a specified threshold. The task \texttt{crossdriz} was used on the cleaned images to create a set of cross-correlation frames. Then the shifts between images were determined using the cross-correlation frames, and images were combined with the drizzle algorithm (Fruchter et al. \citealp{Fruchter}) using a bad pixel mask created earlier.

\subsection{Photometry}
\

\begin{figure*}

\centering $
\begin{array}{cc}
\includegraphics[trim=10mm 5mm 5mm 8mm, clip,width=70mm]{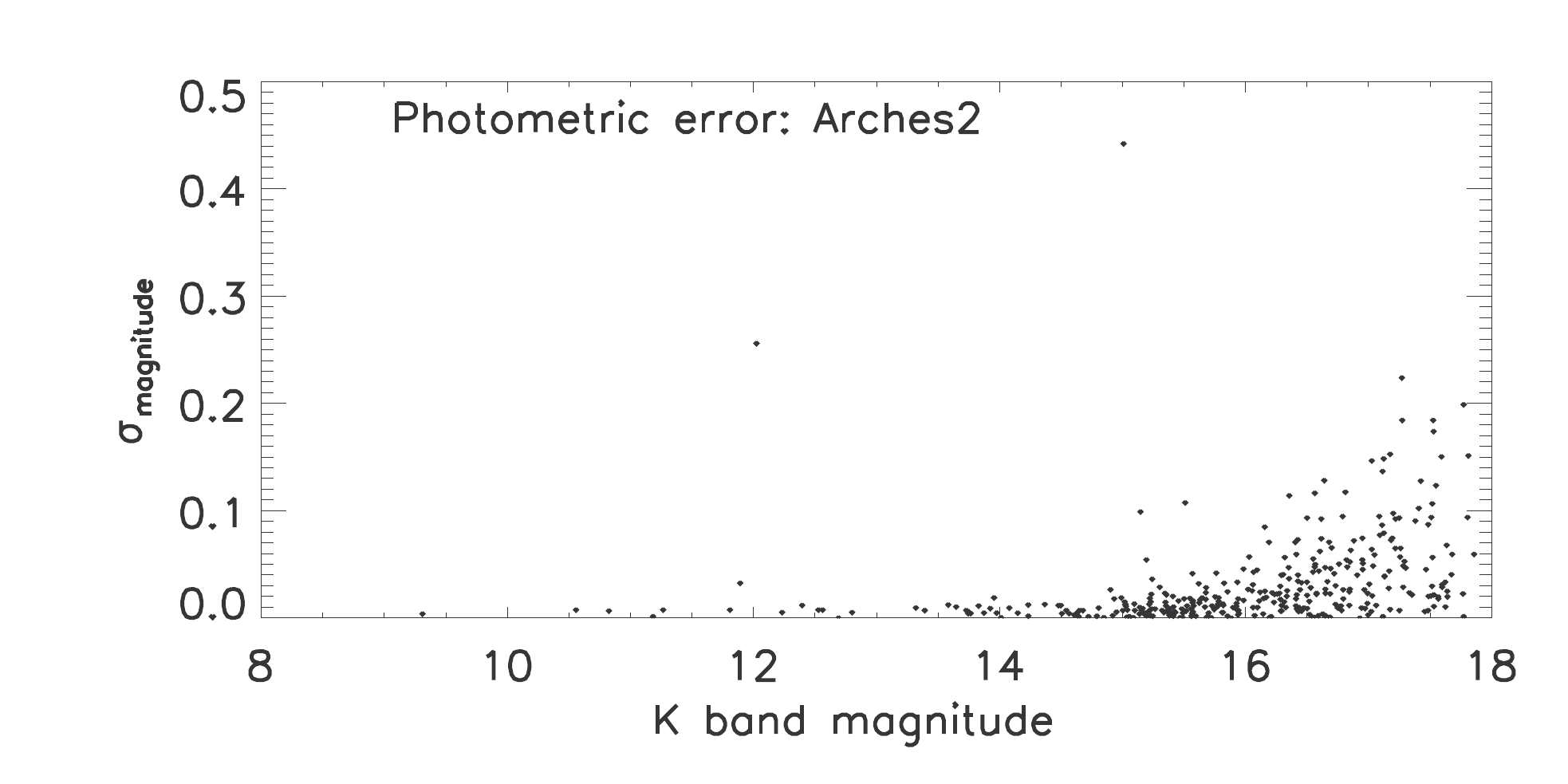} &
\includegraphics[trim=10mm 5mm 5mm 5mm, clip,width=70mm]{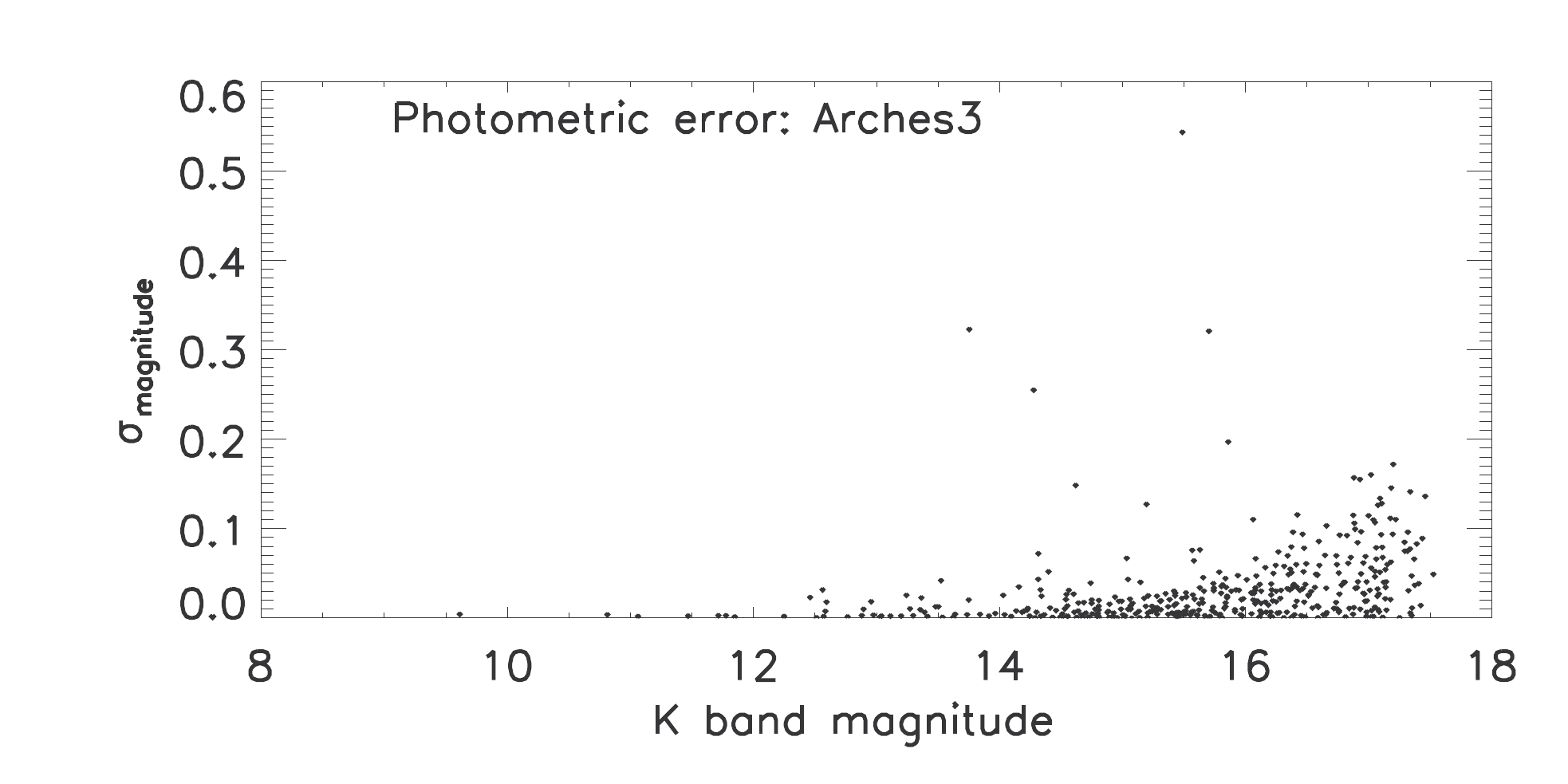} \\
\includegraphics[trim=10mm 5mm 5mm 8mm, clip,width=70mm]{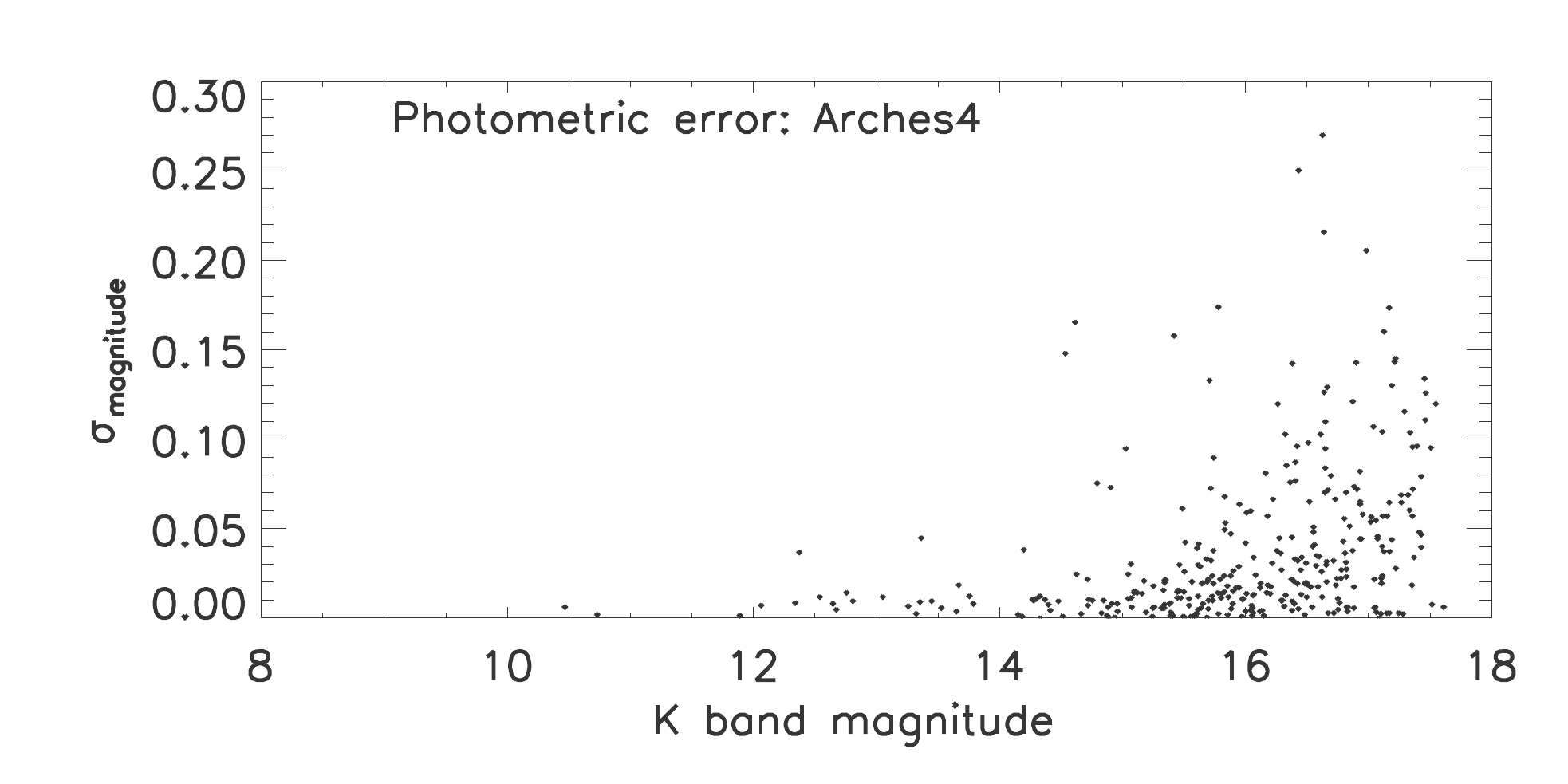} &
\includegraphics[trim=10mm 5mm 5mm 8mm, clip,width=70mm]{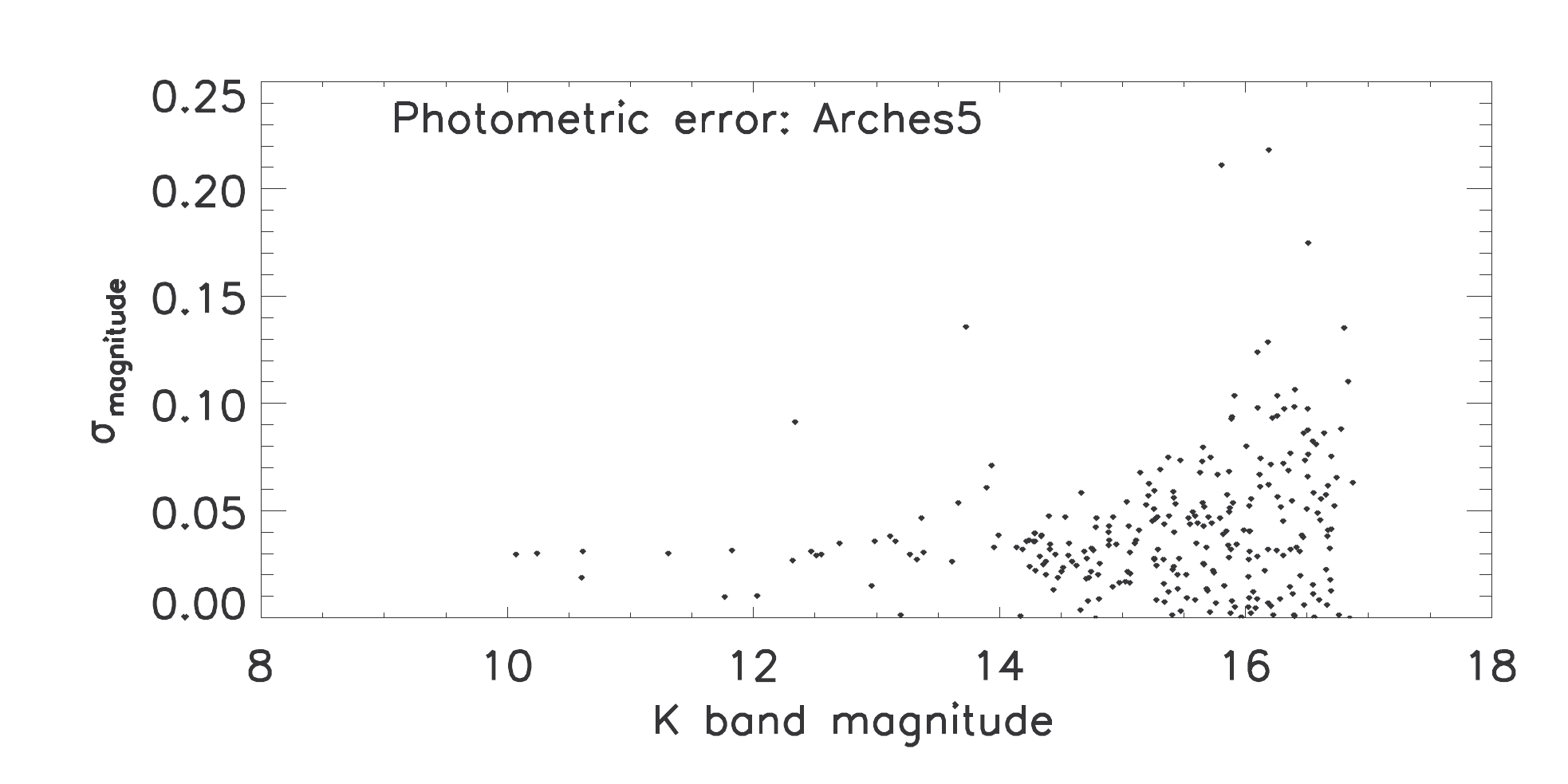} 
\end{array}$
\caption{Standard deviations of the individual $K_{s}$-band magnitudes in the three auxiliary frames. As expected, the brighter sources have smaller photometric uncertainties. The value of $\sigma_{magnitude}$ for the majority of the sources is less than 0.1 mag.}
\label{phot_err}

\end{figure*}

The photometry was extracted using Starfinder, an IDL PSF-fitting package specifically developed to analyze AO data (Diolaiti et al.~\citealt{Diolaiti}). 
Starfinder's PSF-fitting algorithm extracts the empirical PSF from an image by a median combination of suitable stars. For this purpose we selected 10 to 15 reference stars in each field, chosen to be relatively isolated, bright, and equally distributed across the field. However, since it is not easy to find such stars in crowded fields, secondary sources  close to selected stars were first removed. A circular mask of 80 pixel diameter was used as the PSF size. Starfinder performs a cross-correlation check to determine the similarity of potential stars with the PSF. For the source extraction we set the correlation threshold to $\geq 0.7$ to accept sources as potential stars. The final PSF was extracted after two cleaning iterations, and three iterations were performed to detect stars with a count level of 3$\sigma$ above the average background in the image.

Starfinder assumes a constant PSF across the image, which is not exactly the case in wide-field AO imaging. That AO performance is highly dependent on the distance from the guide star (anisoplanatism) can lead to a spatial trend in the amplitude and shape of the photometric residual in the PSF subtracted frames. We used 10-15 stars in each field to extract the average PSF for PSF-fitting. To account for the variation in the residual flux, $F_{res}$, the ratio of the flux in the PSF-fitting residual to the stellar flux, $\frac{F_{res}}{F_{star}}$, was plotted against position. We detected a systematic linear trend with position, which could affect our photometry by 0.2 mag in fields 2, 3, and 5. We corrected the stellar fluxes, $F_{star}$, by fitting a linear function, ${\cal C_{\mathrm{fit}}}(x)$ to the flux ratio, $\frac{F_{res}}{F_{star}}$:

\begin{equation}
\label{e_corr_factor_error}
F_{\mathrm{corr},star}(i) = F_{star}(i)+({\cal C_{\mathrm{fit}}}(x) \times F_{star}(i)) \,
\end{equation}
where $x$ is the position of the $i$th star. This flux correction allows a uniform zeropoint correction to be applied across each field (see Sect. \ref{sec:calib}).

To measure the photometric errors, we created three sublists of images for each field, each containing an equal number of images in such  a way that images with similar qualities (Strehl ratio) are distributed equally in the lists. The three lists were drizzled into three auxiliary frames for each field. We then applied the same Starfinder source extraction as in the deep science images.  We estimated the photometric error of each star by measuring the standard deviation of the independent magnitude measurements in the three auxiliary images (Fig. \ref{phot_err}). The value of $\sigma_{magnitude}$ for the majority of the sources is less than 0.1 mag.

\subsection{Calibration}\label{sec:calib}

Since there are no red standard stars available, we use local standards in our fields for zeropoint calibration employing the NACO ground-based photometry of Espinoza et al. \cite{espinoza}. Espinoza et al. \cite{espinoza} work in the natural photometric  system of  NACO in order to avoid large color terms from unreddened reference stars and use NACO standard stars observed in the same night for their absolute calibration. 

We used  their calibrated sources in the $JHK_{s}$ photomotry list (their Table 3) located in the inner 0.4 pc from the cluster center for zeropointing. The wide $J$-band frame was calibrated based on 38 local standards in the central 0.4 pc of the Arches cluster. The only available $H$-band data in the central field (Fig. \ref{fields}) was calibrated utilizing 117 stars in the central region (Field 1). The $K_{s}$ calibration for the central field was performed by the direct comparison of 125 stars in Field 1. To calibrate the outer fields, we used two $K_{s}$ calibration frames obtained in visitor mode in June 2012. These two short exposure calibration fields (see Table \ref{ob_log}) centered on the boundaries of our original fields allowed us to determine the zeropoints with respect to Espinoza et al.~\cite{espinoza} and 
our calibrated central field (see Fig. \ref{fields}). Fields 3, 4, and 5 were thereafter independently calibrated with respect to the calibration fields based on 59-111 stars (see Table \ref{calib_info}).

Since we were not able to obtain a calibration reference for Field 2, and this field also has no overlap region with the remaining data set, the zeropoint for Field 2 was determined
using the UKIDSS database. This survey is resampled to a uniform spatial resolution of $1"$, which is improved by a factor of two  compared to 2MASS. UKIDSS also probes three magnitudes deeper than 2MASS, such that it provides more accurate zeropointing, especially in crowded cluster fields (Lucas et al. \citealp{Gps}). The low resolution allowed for only four calibration stars, such that we consider the $K_s$-band calibration of Field 2 more uncertain than the zeropoints of the remaining fields.

\begin{table*}
\caption{Calibration information}\label{t_observations}
\centering
\begin{tabular}{cccccc} 
\hline\hline             
Field & Filter & Calib. Ref. & $N_{matched}$ & Err.$^{\mathrm{(a)}}$ & residual rms.$^{\mathrm{(b)}}$  \\

\hline
  \multicolumn{6}{c}{\it science frames }\\
\hline
wide field  & J	 & Espinoza et al.~2009         &	36	& 0.011 & 0.066	 \\
1           & H	 & Espinoza et al.~2009         &	117	& 0.007 & 0.078  \\
1           & Ks & Espinoza et al.~2009         &	125	& 0.005 & 0.056	 \\
2           & Ks & Lucas et al.~2008 (UKIDSS/GPS) &	 4 	& 0.049 & 0.024	 \\
3           & Ks &          calib3-5            &	111	&  0.017    & 	 0.184   \\
4           & Ks &          calib4-5            &	59	& 0.011 & 0.086   \\
5           & Ks &          calib3-5            &	98	& 0.012 &  0.120	 \\

\hline
  \multicolumn{6}{c}{\it calibration frames} \\
\hline
calib3-5            & Ks & Espinoza et al.~2009 & 86		&  0.097     & 0.010\\
calib4-5            & Ks & calibrated F1 & 45	&    0.019   & 0.129	  \\

\hline
\end{tabular}

\label{calib_info}
\tablefoot{
  \tablefoottext{a}{Zeropoint error $=$ standard deviation $/ \sqrt{N}$ } 
  \tablefoottext{b}{Residual rms $=$ standard deviation of the sources used for zeropointing. }
  }
\end{table*}

\section{Extinction derivation}
\label{sec:exti}

The extinction of each star can be derived by individual dereddening in the color-magnitude plane. In this section, a sample of cluster members is selected from the color-magnitude diagram (CMD) of each field to determine the individual extinction for each cluster star candidate. 

\begin{figure}[!ht]
\centering 

\includegraphics[trim=1mm 0mm 0mm 0mm, clip,width=3in]{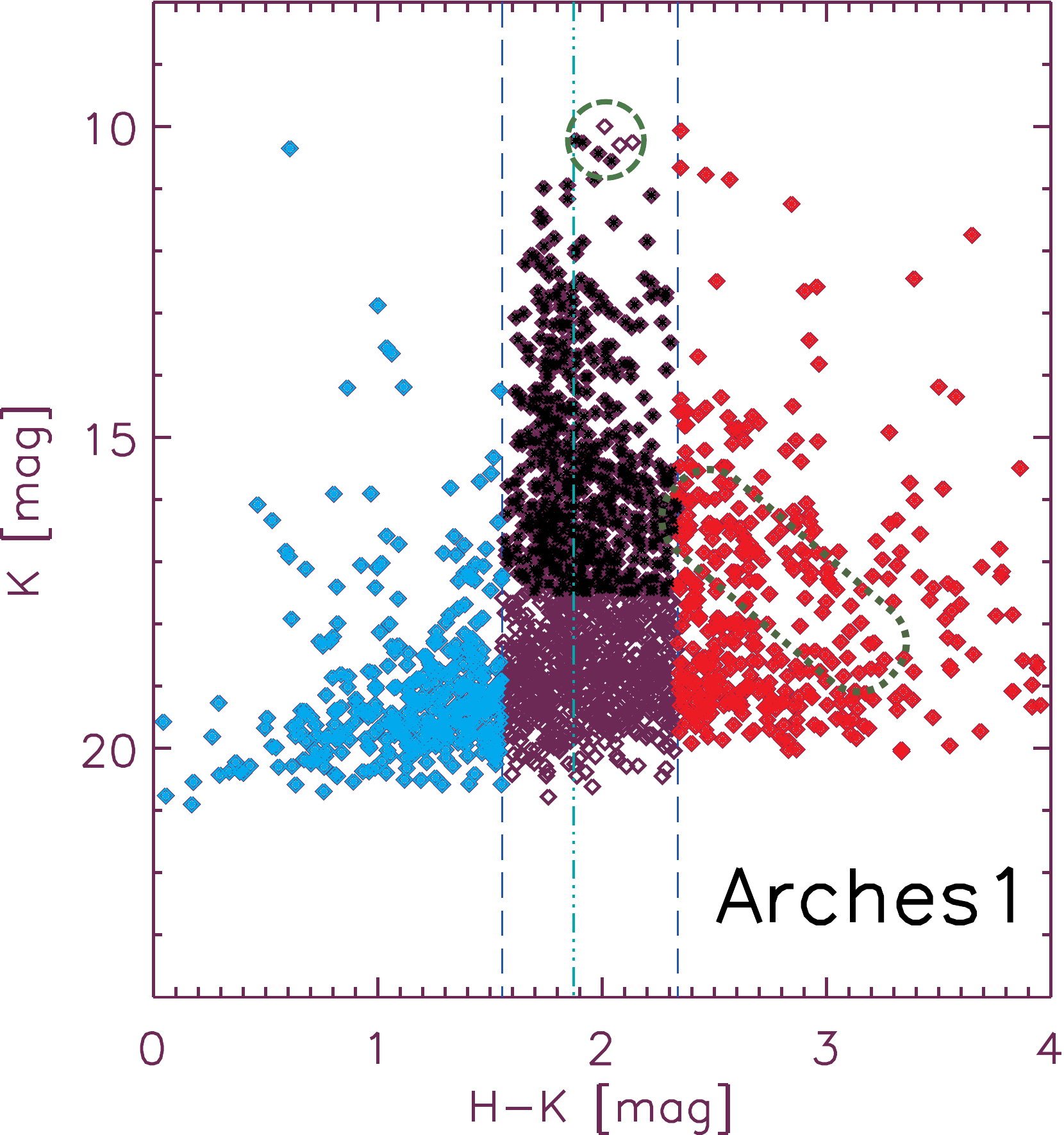} 

\caption{Color-magnitude diagram of the center of the Arches cluster.  Vertical dashed lines show the color selection used to discard contaminating background and foreground sources. Less extincted foreground sources, mostly associated with the spiral arms, are shown by blue diamonds, whereas intrinsically red objects, such as red giants or red clump stars are marked in red. The location of red clump stars is indicated with an enclosed dotted line. The black diamonds  represent likely cluster members. The mean H-$K_{s}$ color of the bright cluster members, $11< K_{s} <15$ mag, in the center of the Arches cluster is shown by  a green dot-dashed line. Brighter sources, $K_{s} < 11$ mag, belong to the known WR population in the Arches cluster (denoted by the enclosed dashed circle).}
\label{CMD1}
\end{figure}

\begin{figure*}[!ht]
\centering $
\begin{array}{cc}
\includegraphics[trim=21mm 5mm 10mm 12mm, clip,width=2in]{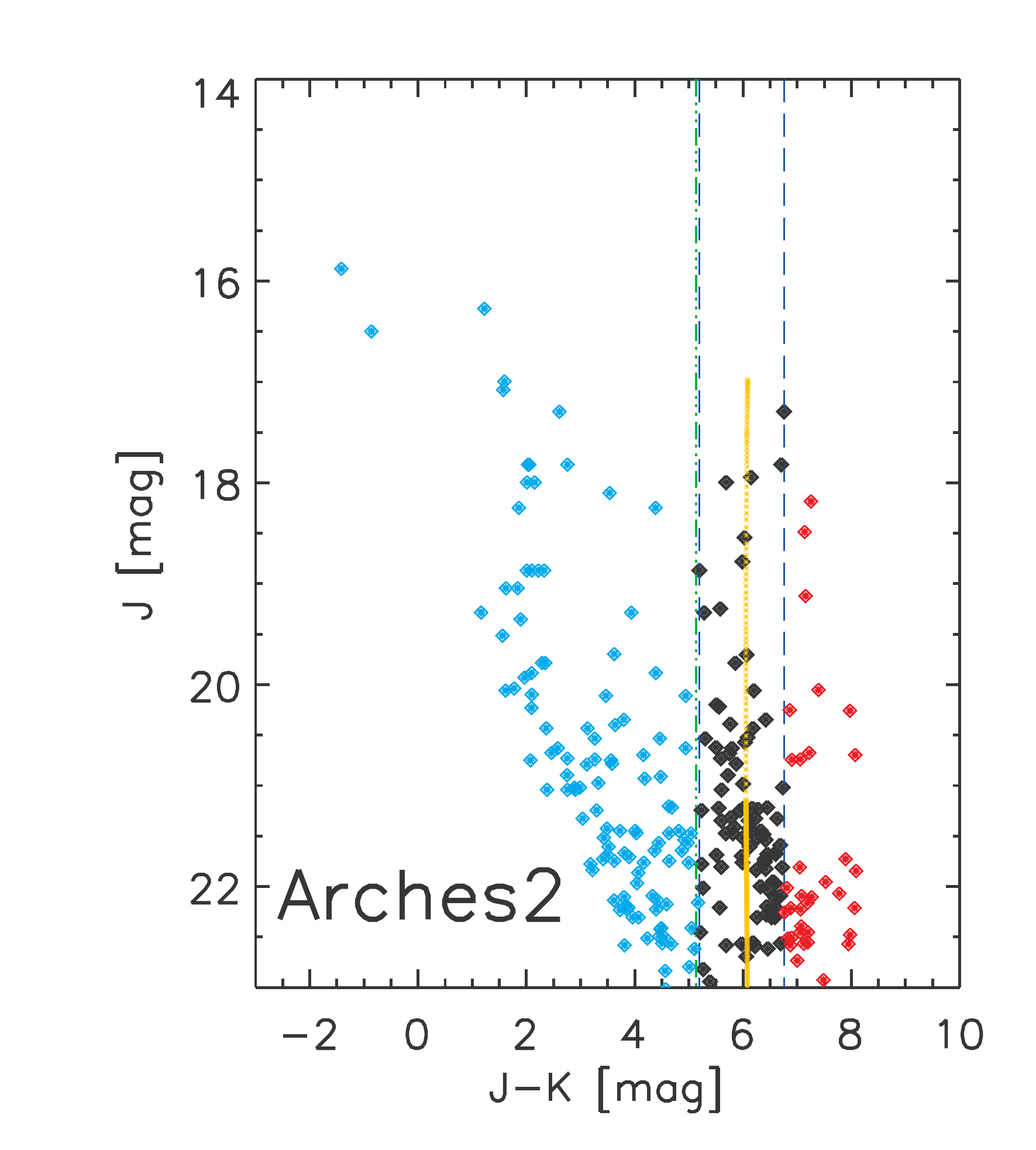} &
\includegraphics[trim=21mm 5mm 10mm 12mm, clip,width=2in]{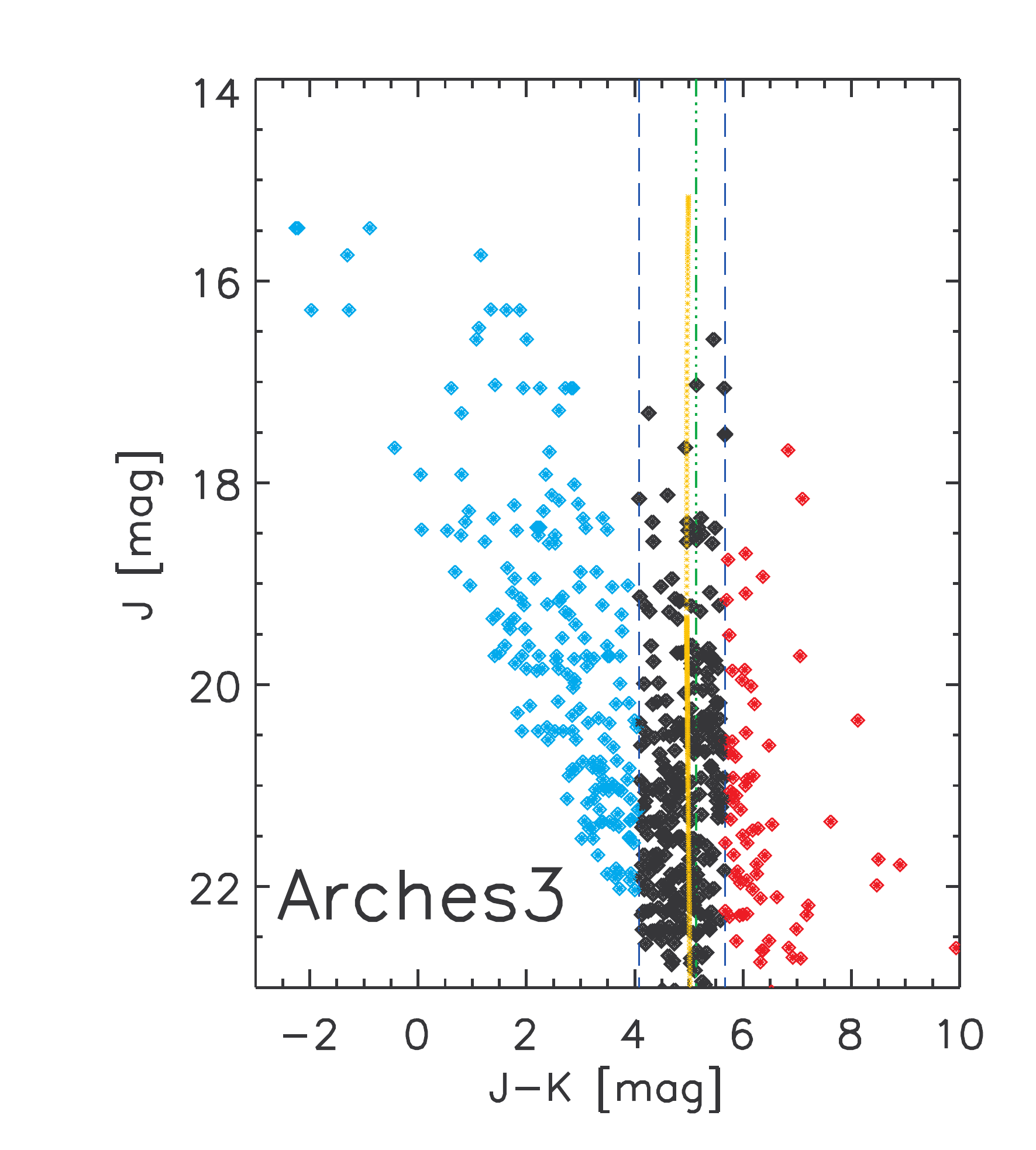} \\
\includegraphics[trim=21mm 5mm 10mm 12mm, clip,width=2in]{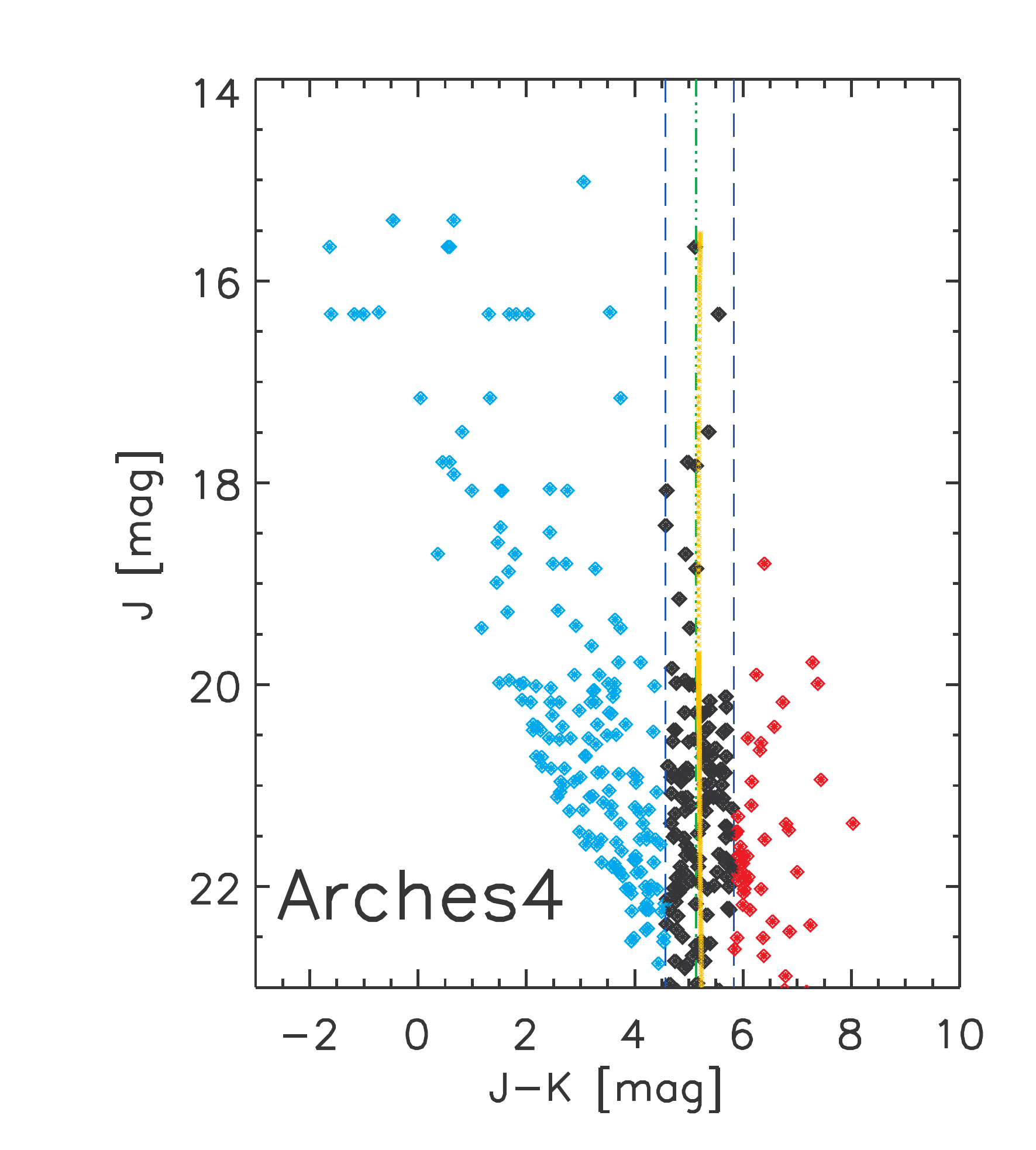} &
\includegraphics[trim=21mm 5mm 10mm 12mm, clip,width=2in]{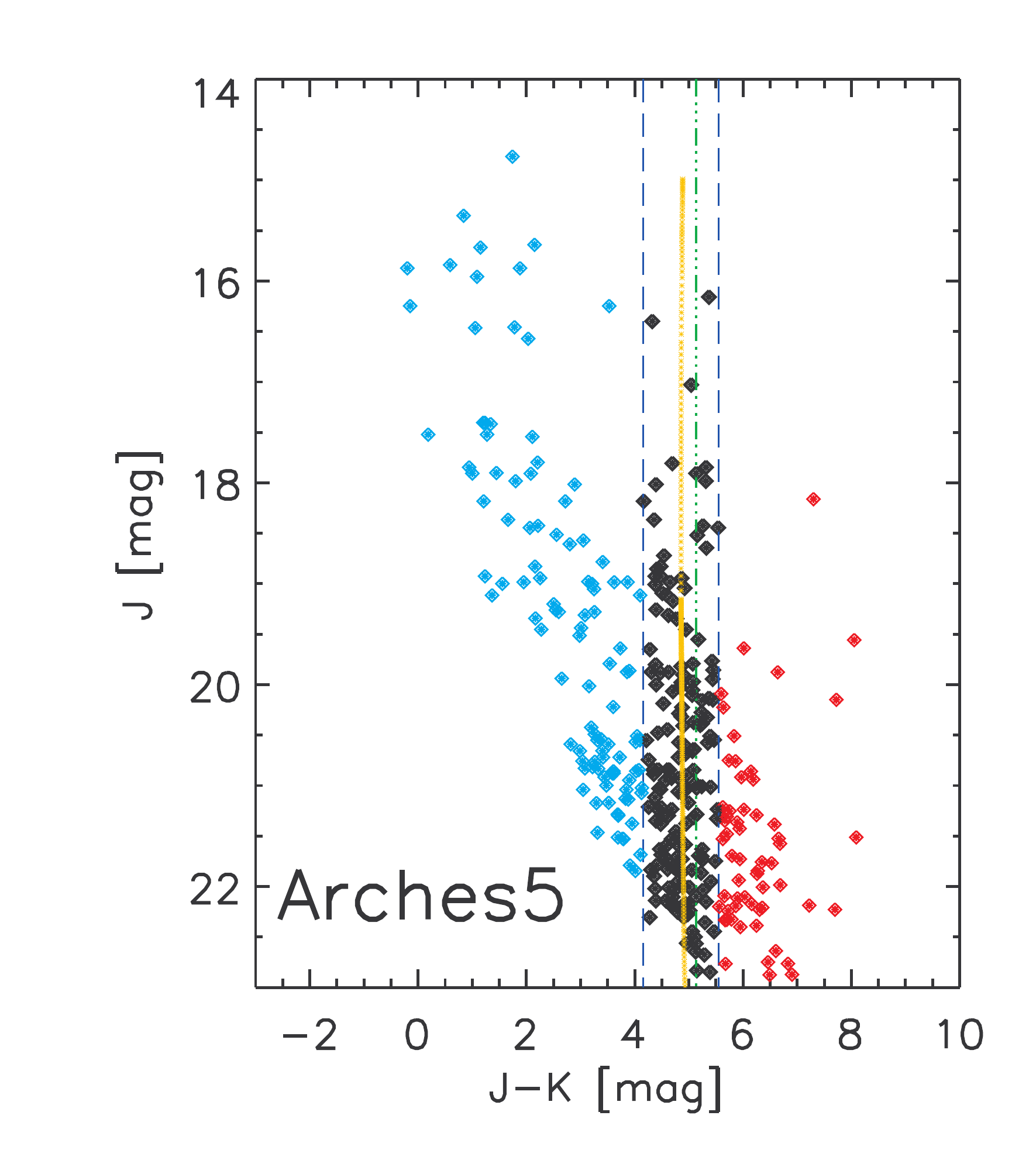} 
\end{array}$
\caption{ Color-magnitude diagrams of the outskirts  of the Arches cluster. Blue dashed lines show the color selection used to choose the cluster members and discard  background (blue diamonds) and foreground (red diamonds) sources. The Geneva isochrone with the mean extinction value of the cluster members in each field is shown in yellow. The green dot-dashed line represents the approximate mean color of the cluster members in the center of the cluster where the mean extinction is slightly lower.}

\label{CMDs}
\end{figure*}

The CMD for the central field is shown in Fig.~\ref{CMD1}. A population of blue foreground sources with lower extinction and mostly associated with the spiral arms,  as well as intrinsically red objects such as red giants or red clump stars, can be distinguished in Fig.~\ref{CMD1}. We applied a color selection (Fig.~\ref{CMD1})  in order to reject contaminating sources and obtain a representative sample of cluster members. The color selection is based on the prominent main sequence cluster members in the CMD of the central part of the Arches cluster. The bright part of the observed sequence in the central field, $11< K_{s} <15$ mag, has a mean color of $A_{H}-A_{K_{s}} \sim H-K_{s} = 1.87$ mag (Fig. \ref{CMD1}), where we have assumed the intrinsic $H-K_s$ color of young, high-mass main sequence stars to be close to zero ($-0.03 < H-K_s < -0.04$ (Lejeune \& Schaerer \citealt{geneva}). This color can be converted to an approximate color of $J-K_{s} \sim 5.1$ mag (using Rieke \& Lebofsky 1985). Brighter sources with $K_{s} <11$ are likely to  belong to the known WR population in the Arches cluster. Using the expected main sequence color of cluster stars from Field 1 to identify the cluster main sequence in each outer field, we determine appropriate color selections for likely cluster members (Fig.~\ref{CMDs}). The mean extinction in each outer field varies slightly with respect to Field 1. In Fig.~\ref{CMDs}, we show the expected $J-K_s$ main sequence color derived from the mean $H-K_s$ color in Field 1. The higher extinction in Field 2, which has the largest radial distance from the cluster center, is particularly pronounced. An increase in extinction with increasing radius is consistent with the previously detected radial extinction variation (Stolte et al. \citealt{stolte_2002}; Espinoza et al. \citealt{espinoza}).

The age of the Arches cluster is estimated to be 2.5 Myr (Blum et al. \citealt{blum}; Najarro et al. \citealt{Najarro}; Martins et al. \citealt{martins}, for a detailed discussion of the previous studies on the age and the metallicity of the cluster see Sect. \ref{error-PDMF}).
 We adopted a Geneva isochrone  (Lejeune \& Schaerer \citealt{geneva}) with solar metallicity  located  at the GC distance of 8 kpc (Ghez et al. \citealt{Ghez2008}).

To derive the individual extinction values we use two extinction laws (EL): Rieke \& Lebofsky  \cite{rieke1985} and Nishiyama et al. \cite{Nishi2009}. As mentioned in the introduction, the extinction law derived toward the GC by Rieke \& Lebofsky \cite{rieke1985} ($A_{\lambda}\propto\lambda^{-1.61}$ which implies $\frac{A_{j}} { A_{K}}=2.51,\frac{A_{H}}{A_{K} }= 1.56 $) has been used in most stellar population studies in the GC until recently. Nishiyama et al. (2009) derived the extinction law again toward the GC and obtained  a power law of steeper decrease with wavelength ($A_{\lambda}\propto\lambda^{-2.0}$ which implies $\frac{A_{j}} { A_{K_{s}}}=3.02,\frac{A_{H}}{A_{K_{s}} }=1.73 $), which is also consistent with the recent determination of the near-infrared extinction law by Schoedel  et al. \cite{schoedel} and Stead \& Hoare \cite{Stead} along the GC line of sight.

\begin{figure}[HT]
 \centering
\includegraphics[trim=22mm 5mm 5mm 12mm, clip,scale=0.5]{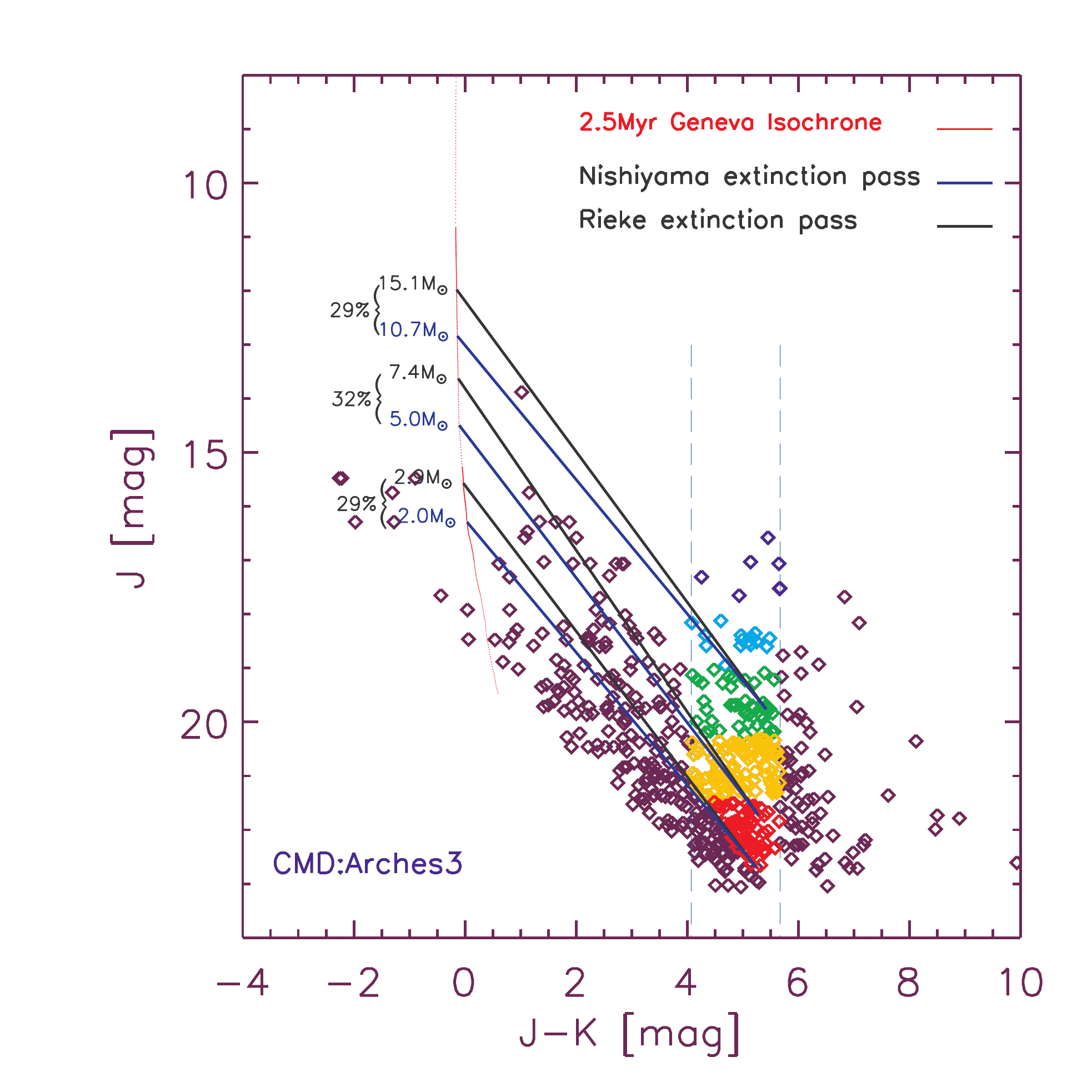}
\caption{The color-magnitude diagram of Field 3 in the outskirts  of the Arches cluster. Cluster members are selected between the two dashed lines and are color coded according to their $J$-band magnitude. A Geneva isochrone of solar metallicity with an age of 2.5 Myr is also shown. Black lines represent the extinction path assuming the Rieke \& Lebofsky  \cite{rieke1985} extinction law,  while blue lines are extinction paths based on the Nishiyama et al. \cite{Nishi2009} extinction law. The  difference in derived masses using the two laws are written for sample sources close to the isochrone in percentage. }
\label{ilus-cmd}
\end{figure}

Using two ELs translates into two extinction path slopes in the CMD (Fig. \ref{ilus-cmd}). We slide back the cluster members along each extinction path toward the nonextincted theoretical isochrone. The brightness, color and mass of a star at the intersection point of the unreddened isochrone with the extinction vector are assumed to be the  intrinsic brightness, color, and mass of this star (Fig. \ref{ilus-cmd}). Three of the brightest sources in Field 1 are formally located above the maximum initial isochrone mass of 120$M_{\bigodot}$ after dereddening. These sources are discarded since we are not able to derive their intrinsic properties. In this work we do not consider binarity  or rotation of the detected sources, which can also partly contribute to the observed color of each star.

\subsection{Comparison of the parameters derived from the two extinction laws}\label{sec:comp}

\begin{figure*}[h]
\centering $
\begin{array}{cc}
\includegraphics[trim=18mm 5mm 7mm 12mm, clip,scale=0.5]{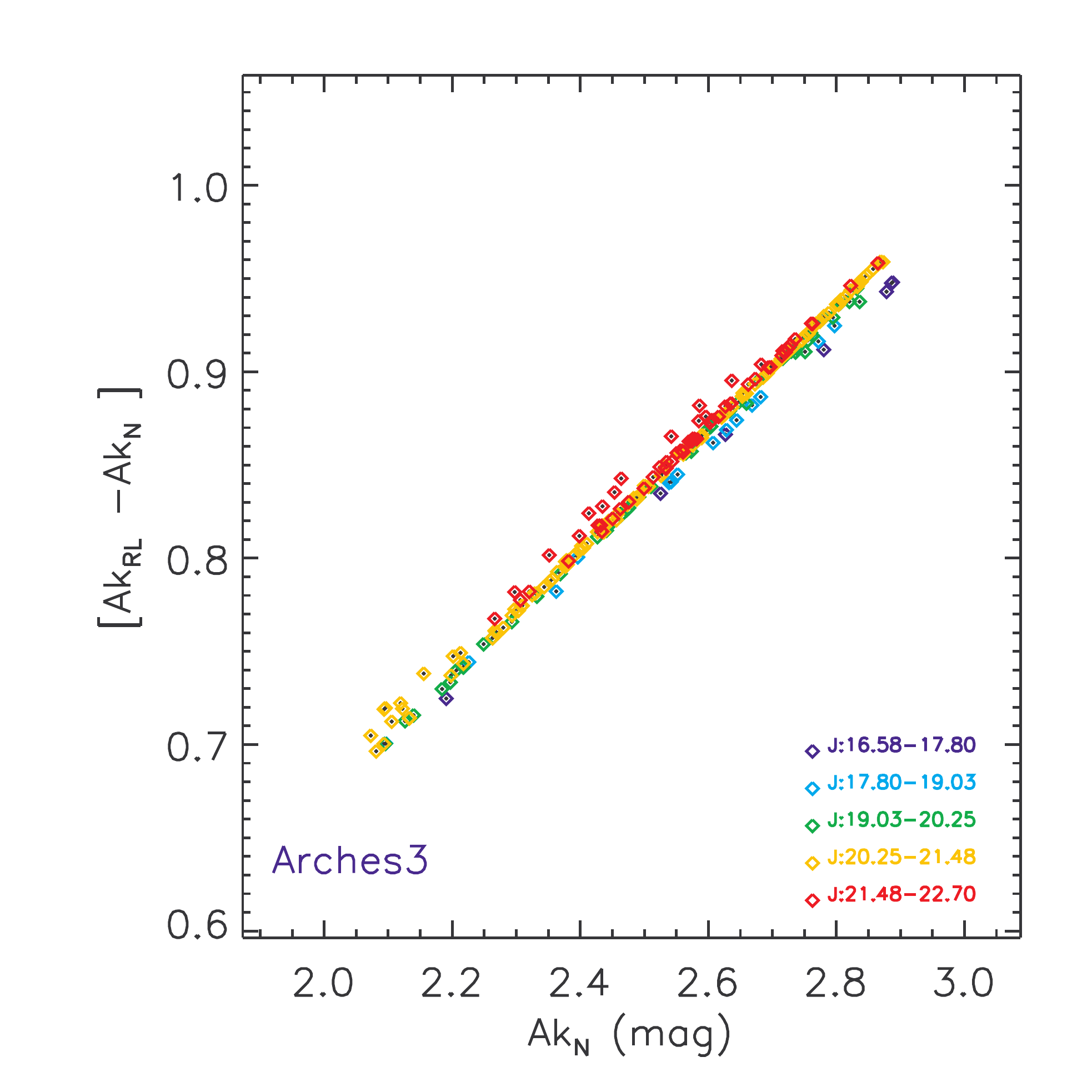} &
\includegraphics[scale=0.5]{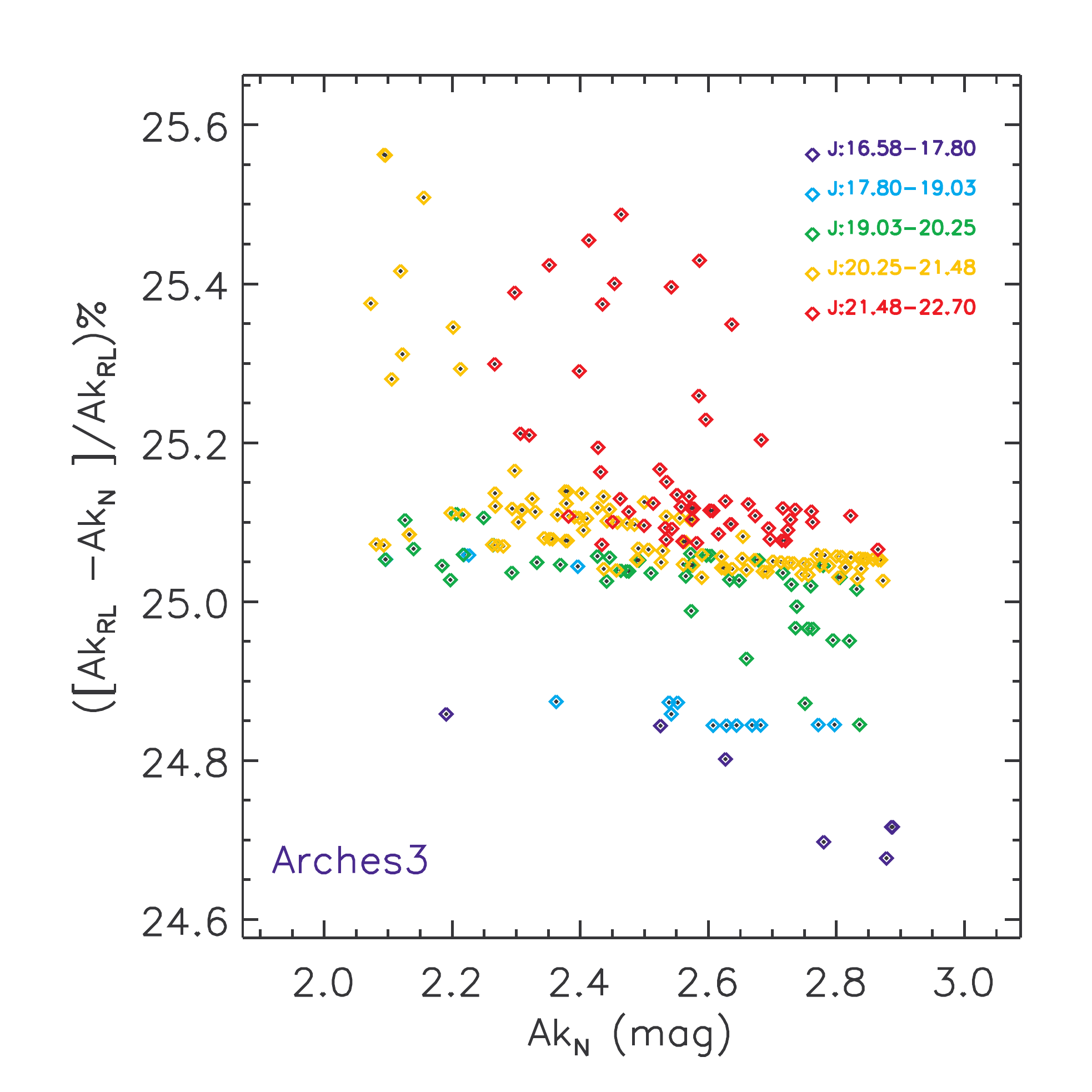} \\
\end{array}$
\caption{{\sl Left:} The extinction difference in Field 3 (outskirts of the cluster) derived using the Nishiyama et al. \cite{Nishi2009} and the Rieke \& Lebofsky \cite{rieke1985} extinction laws is shown as a function of the $K_{s}$-band extinction derived assuming the Nishiyama et al. extinction law. The difference depends linearly on the extinction and increases with increasing extinction. {\sl Right:} The difference in percentage. The  average $A_{K_{s}}$ difference is $\sim 25\%$ in Field 3 and $\sim 24\%$ across the whole cluster.}
\label{exti_dif}
\end{figure*}

Extinction causes stars to be reddened with respect to their intrinsic color. Assuming different extinction laws will change the derived extinction and consequently the intrinsic properties of the stars. Figure \ref{exti_dif} (left) strikingly illustrates the $K_{s}$-band extinction difference, $\Delta A_{K_{s}}$, for each star in Field 3 using the Rieke \& Lebofsky \cite{rieke1985} EL (RL-EL)  and the Nishiyama et al. \cite{Nishi2009} EL (N-EL). The $A_{K_{s}}$ difference across the whole cluster is more than 0.6 mag, and it can reach up to 1.1 mag which is equivalent to roughly ten magnitudes of visual extinction. The difference increases linearly with increasing extinction and is highest for the sources with the highest $A_{Ks}$ values.  The  average $A_{K_{s}}$ difference in the cluster is $\sim 24\%$. The $A_{K_{s}}$ difference in percentage (right) and absolute value (left) for Field 3 is shown in Figure \ref{exti_dif}. In Field 3, most of the sources have  an $A_{K_{s}}$ difference of $\sim 25\%$ while some fainter sources appear in a small upward spread in this plot. The less certain photometry of fainter sources, together with a small curvature at the low-mass end of the isochrone, causes this spread: since the length of the connecting path from a star to the isochrone is the measure of the stars' extinction, slightly different slopes of the connecting path result in a bigger difference in the length of the line in areas where the ischrone has a small bend.

\begin{figure}[h]
\begin{center}

\includegraphics[trim=19mm 5mm 8mm 10mm, clip,scale=0.5]{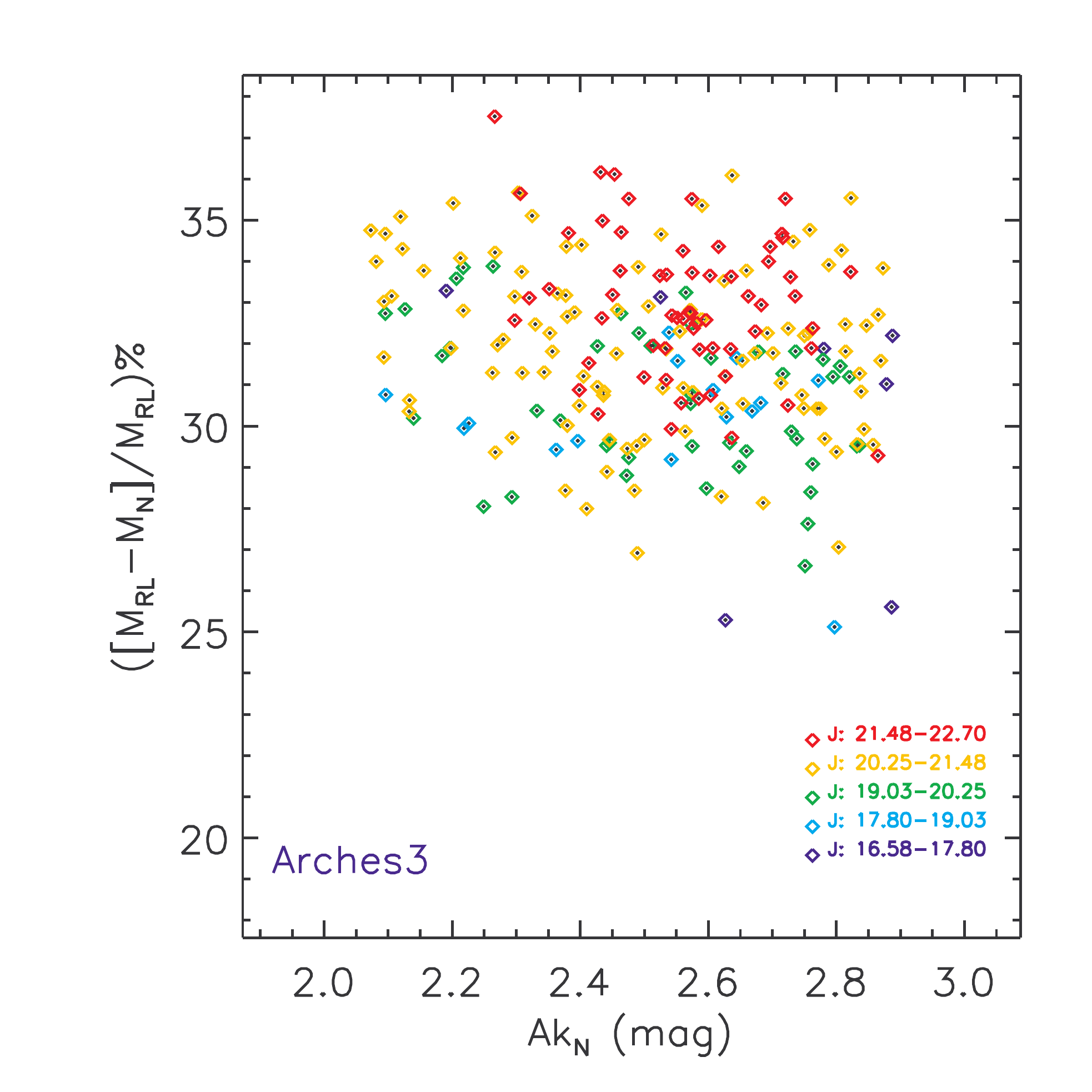}
\caption{The derived mass difference assuming the Nishiyama et al. 
\cite{Nishi2009} and Rieke \& Lebofsky \cite{rieke1985} extinction laws is shown as a function of $A_{K_{s}}$ for each source in Field 3. Derived masses using the Rieke \& Lebofsky extinction law are on average 30\% higher than derived masses when using the Nishiyama et al. extinction law.}
\label{mass-dif}
\end{center}
\end{figure}

Figure \ref{mass-dif} shows the difference in the derived initial mass  for each star using the two different ELs for Field 3. Derived masses using the RL-EL are $\sim$ 30\% higher than derived masses assuming the N-EL for the full sample of Arches cluster members. While the most massive initial mass in our sample is 104$M_{\bigodot}$ when dereddening with an RL-EL, the highest initial mass is only 80$M_{\bigodot}$ when the N-EL is used. In Fig. \ref{mass-dif}, the less certain photometry of faint sources and a small bend in the faint end of the isochrone act against the more rapid increase in masses at the bright end of the isochrone. Therefore in Fig. \ref{mass-dif} there is a vertical spread of a few percent among both faint and bright sources. In their study, Kim et al. \cite{kim2006} briefly report that the 50\% completeness limit of $1.3 M_{\bigodot}$ derived from the Rieke, Rieke \& Paul \cite{rrp} extinction law decreases to $1 M_{\bigodot}$ when applying Nishiyama et al. \cite{Nishi2006} instead. This comment is consistent with our finding that the individual stellar mass decreases by $\sim 30$\%
when the Nishiyama extinction law is employed.

\begin{figure}[!ht]
\centering 

\includegraphics[trim=22mm 5mm 5mm 12mm, clip,scale=0.5]{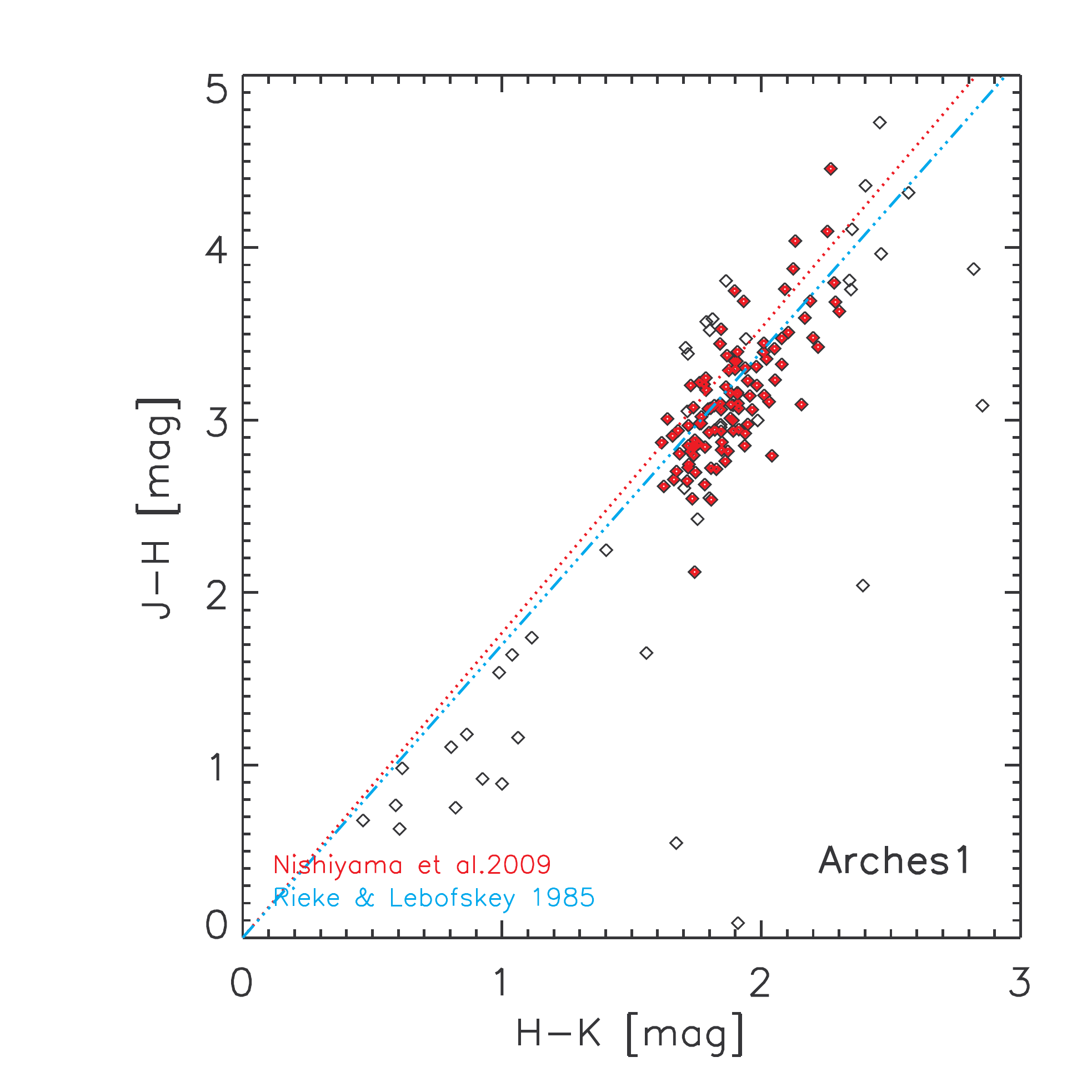} 

\caption{Color-color digram for the central field (Field 1), using the Subaru $J$-band magnitudes, together with NACO high-resolution $H$ and $K_s$ magnitudes, to test whether conclusions on the best-fitting extinction law can be drawn from these data sets.}
\label{CCD_f1}
\end{figure}

Figure \ref{CCD_f1} illustrates the color-color digram for the central field (Field 1). Our $J$-band data for the central region, the only field where we have additional NACO $H$-band observations, is not employed to derive stellar properties since seeing-limited $J$-band data will perform less efficiently in the crowded central region. Nevertheless, we used the Subaru $J$-band magnitudes to test whether any conclusions on the best-fitting extinction law can be drawn from these data sets. The present spread in Fig. \ref{CCD_f1} shows that both extinction law slopes are consistent with our data within the photometric uncertainties. We repeated the same experiment with the data from Espinoza et al. \citep{espinoza} since their work benefits from the $J$-band AO images acquired by the NACO AO system on the VLT. However, scatter in their data also does not allow us to distinguish between the two extinction paths.

\subsection{Extinction map}
From the individual extinction values we can construct the extinction map. Ideally, we need to have the extinction value for  every point in our field. Since the extinction is only known for the places in which we observe stars, creating the extinction map means assigning each star's extinction value to its neighborhood. We use Voronoi diagrams to define the neighborhood of each star on the plane of the sky. Considering a 2-D plane ($\mathbb R^2$) with a finite data set of $n$ sites (stars) $S=\lbrace s_1,s_2,...,s_n\rbrace $, the $i$-th Voronoi cell consists of all points $( x \in \mathbb R^2)$ whose distance to its generating site, $s_{i}$, is not greater than to any other site in the plane, $V(s_i)=\{ x \in \mathbb R^2 \mid \forall j\neq i,d(s_i,x)< d(s_j,x)\}$. 
 
   \begin{figure*}
   \centering
   \includegraphics[scale=3]{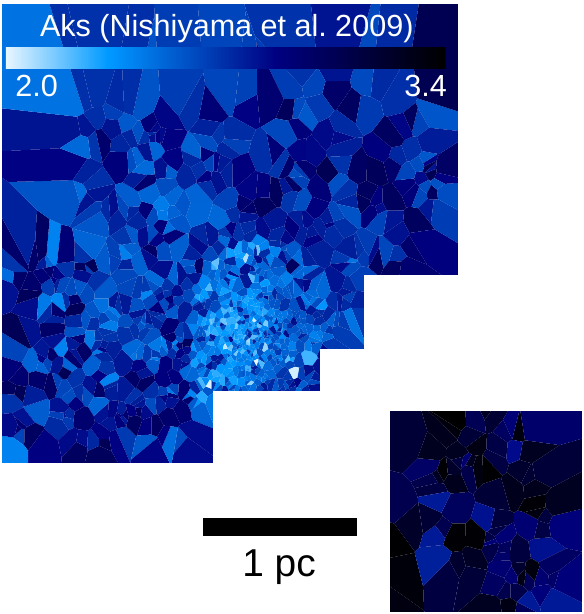}
   \caption{The extinction map of the Arches cluster using Voronoi diagrams. Each star is associated with one and only one cell, the color of which is determined by the measured extinction value at the location of the star. A region of lower extinction is present in the center of the cluster, while stripes of higher extinction are present in the southwest and partly northwest of the cluster. The extinction is high and varies by up to $\sim 2$ mag across the cluster. The extinction values derived based on the Nishiyama et al. \cite{Nishi2009} extinction law  vary between $2<A_{K_{s}}<3.4$ mag, while utilizing the Rieke \& Lebofsky \cite{rieke1985} extinction law yields an extinction range of $2.7<A_{K_{s}}<4.5$ mag (not shown). The structure of the two extinction maps based on the two different extinction laws is only marginally different. North is up and east to the left.}
\label{exti_map}
\end{figure*}

The Voronoi diagram results in a polygon partition of the plane. Overdensities are represented as regions of small area and homogeneously spread points are reflected as  polygons with comparable area, while a preferred orientation in the distribution of the  points is exhibited as oriented shapes (Aurenhammer \& Klein  \citealt{voronoi}). The resulting extinction map derived assuming the N-EL is shown in Fig.~\ref{exti_map}. Bright colors correspond to regions of low extinction while dark patches represent high-extinction areas. The inverse of the area of each Voronoi cell shows the precession of the extrapolation of the star's extinction to its  region. As discussed in Sect.~\ref{sec:comp}, using different extinction laws would alter the individual extinction value between 0.6 and 1.1 mag (see Fig.~\ref{exti_dif}). However, the structure of the extinction map remains very similar when applying different extinction laws since differences in the spatial extinction distribution are small.

The $K_{s}$-band extinction derived employing the RL-EL varies between $2.7 <A_{K_{s}}<4.5$ mag. In the inner core of the cluster, we see an average of $\langle {A_{K_{s}}} ({r<0.2}$ pc$)\rangle=3.3 \pm 0.3$ mag, as compared to a mean extinction of $\langle {A_{K_{s}}} ({0.2<r<0.4}$ pc$)\rangle=3.4 \pm 0.3$ mag in the intermediate annulus and $\langle A_{K_{s}} ({r>0.4}$ pc$)\rangle=3.5\pm 0.4$ mag in the cluster outskirts. The highest extinction is found in the southwest of the cluster (Field 2) while the least extinction is detected toward the center (Field 1). The  acquired extinction values applying the N-EL are on average 0.8 mag lower than values derived based on the RL-EL. The extinction derived from the N-EL varies within the range of  $2<A_{K_{s}}<3.4$ mag. The mean extinction values across the inner region of $r<0.2$ pc, the annulus with $0.2 <r<0.4 $ pc, and the region of $<0.4<r< 1.5 $ pc are, respectively,  $2.5 \pm 0.2$, $2.6\pm 0.2$, and $2.6\pm 0.3$ mag. Reported errors on the mean extinction values are standard deviations present among individual extinction measurements in each region.

The derived values of ${A_{K_{s}}}$ using the Rieke \& Lebofsky \cite{rieke1985} EL are consistent with the only previous study with individual extinction correction for the Arches cluster by Espinoza et al. \citep{espinoza}. This study used the  Fitzpatrick \cite{Fitzpatrick}  EL  and either a Bayesian or CMD sliding method to derive extinction values that depends on the brightness and availability of $JHK_{s}$ photometry for each star. They find a $K_{s}$ extinction range of $2.13 < A_{K_{s}} < 4.14 $ mag across a field of view which extends only $\sim 0.5$ pc from the cluster center. They also find a mean extinction value of $\langle A_{K_{s}}\rangle=2.97$ in the inner core ($r<0.2 $ pc) followed by a mean extinction value of $\langle A_{K_{s}}\rangle=3.18$ in the region $0.2 <r<0.4 $ pc. Espinoza et al. \cite{espinoza}  cover the cluster beyond the distance of 0.5 pc partly toward the north and west, where they report a mean extinction of $\langle A_{K_{s}}\rangle=3.24$ for this area (see Espinoza et al. \citealt{espinoza}, Fig. 12). These results are consistent with our findings of  $3.3 \pm 0.3$, $3.4 \pm 0.3$, and $3.5 \pm 0.4$ mag for similar regions based on the Rieke \& Lebofsky \cite{rieke1985} EL, which is comparable to Fitzpatrick \cite{Fitzpatrick}.

 In a recent study, Dong et al. \cite{dong2012} also  calculate individual extinction values  for a number of massive stars in the GC, including 19 sources within a radius of one parsec of the Arches cluster (see Table 2 in Dong  et al. \citealt{dong2012}).  The extinction values are derived using the broad-band filters $J$, $H$ and $K_{s}$ of SIRIUS (Nagayama et al. \citealt{sirius}) and by employing the extinction law of  Nishiyama et al. \cite{Nishi2006}. The mean value of the derived individual $A_{K_{s}}$ extinctions for these 19 sources is 2.53, which is consistent within 0.14 mag with our Nishiyama based mean extinction of  $\langle A_{K_{s}}  (r<1\;pc)\rangle= 2.67 \pm 0.3$.

The region of low extinction in the center of the  cluster is probably due to the presence of massive stars. Massive stars provide strong UV radiation and stellar winds that can disrupt the residual gas and dust (see Stolte et al. \citealt{stolte_2002}). Identified X-ray  sources that coincide with radio emission in the Arches cluster confirm there are powerful ionized winds from late-type Of/Wolf-Rayet stars in the center of the cluster (Lang et al.~\citealt{lang}; Law \& Yusef-Zadeh \citealt{Law-Y}). Regions of higher extinction in the southwest and northwest of the cluster coincide with dark lanes visible in the JHK composite from the UKIDSS GPS survey (Lucas et al.~\citealt{Gps}). Espinoza et al.~\cite{espinoza} also found an area of higher extinction toward the southwest of the cluster, that lies in the gap between Fields 1 and 2 in our work. The presence of relatively low-extinction areas in the eastern part of the outer cluster region is consistent with identified diffuse X-ray emission of the cluster which is elongated towards this area (Law \& Yusef-Zadeh~\citealt{Law-Y}).

Apart from the overall trend of having lower extinction values toward the center of the cluster, we find that the extinction is spatially variable both in the center and at larger radii. The extinction varies  by $\sim 2$ mag in $A_{K_{s}}$  across the cluster ($ \Delta A_{K_{s},RL-EL} =1.8$, $\Delta A_{K_{s},N-EL} =1.4$), which is equivalent to roughly 15-20 mag of visual extinction. Such a high and variable extinction (as shown earlier for the center of the cluster by Espinoza et al. \citealt{espinoza}) implies that using global extinction trends or even single extinction values derived by averaging individual measurements  would alter the results  for the Arches cluster systematically.

\section{Mass function}
\label{sec:mf}

To derive the PDMF of the cluster, we applied a more conservative low-mass selection in addition to the criteria imposed by sensitivity (explained in Sect.~\ref{sec:obs}). We discarded all sources whose masses are less than the mass of the reddest source with a luminosity close to the sensitivity limit. This criterion ensures that the mass function is complete in the faintest mass bin and corresponds to an extinction-limited sample. The lowest mass included in the PDMF is 10-17$M_{\bigodot}$ depending on the field and the extinction law that is applied (see Table \ref{mass_cut_info}). The low-mass truncation is slightly different in the  center since the observational set-up and  completeness limit is different for the center (see Sect.~\ref{sec:obs}). 
Truncating the low-mass end of the mass distribution helps to avoid the field contamination, which is dominated by K and M giants in the Galactic bulge for stars fainter than $J \sim 21$ mag. This is crucial for the outskirts  of the cluster where we expect to have more contamination by field stars relative to the decreasing number of cluster members.

\begin{table}
\caption{The minimum mass of a star that is included in the mass function sample.  }
\label{mass_cut_info}   
\centering               
\begin{tabular}{c c c }        
\hline\hline                 
Field & N-EL & RL-EL  \\    
\hline                        
\\
   min mass: center (F1) & $12.5M_{\bigodot}$ & $17.5M_{\bigodot}$  \\
   
   min mass: outskirts (F2-F5) & $9.7M_{\bigodot}$ & $13.6M_{\bigodot}$  \\
 
\hline                                   
\end{tabular}
\end{table}

\begin{table*}
\caption{Acquired slopes for the mass function derived from the initial and the current masses of the Arches cluster applying the Nishiyama et al. (2009) (N-EL) and Rieke \& Lebofsky (1985) (RL-EL) extinction laws.  The slopes are calculated for the three different regions distinguished by the distance from the cluster center. The last row contains the same slopes for the whole cluster up to a radius of 1.5 pc.
  }             
\label{slopes}      
\centering          
\begin{tabular}{c c c c c c  }     
\hline\hline       
  & \multicolumn{2}{l}{Nishiyama et al. (2009) EL} & &\multicolumn{2}{r}{Rieke \& Lebofsky (1985) EL}\\ 
region (pc) & PDMF, init & PDMF, P-D &  & PDMF, init & PDMF, P-D\\ 

\hline        
            
   $r<0.2$  & -1.63$\pm0.17$ & -1.50$\pm0.35$ & &-1.60$\pm0.17$&-1.20$\pm0.18$ \\  
   $0.2 <r<0.4$ & -2.25$\pm0.27$ & -2.21$\pm0.27$&    & -2.20$\pm0.27$&-2.01$\pm0.19$ \\
   $0.4 <r<1.5$ & -3.16$\pm0.24$ & -3.21$\pm0.30$&     & -3.01$\pm0.19$ &-2.99$\pm0.20$\\
   \hline
   $r<1.5$ & -2.55$\pm0.18$ & -2.53$\pm0.31$ & &-2.43$\pm0.13$ &-2.25$\pm0.14$\\

\hline                  
\end{tabular}
\end{table*}

\begin{figure*}[h]
\centering $
\begin{array}{cc}
\includegraphics[trim=9mm 5mm 0mm 0mm, clip, scale=0.55]{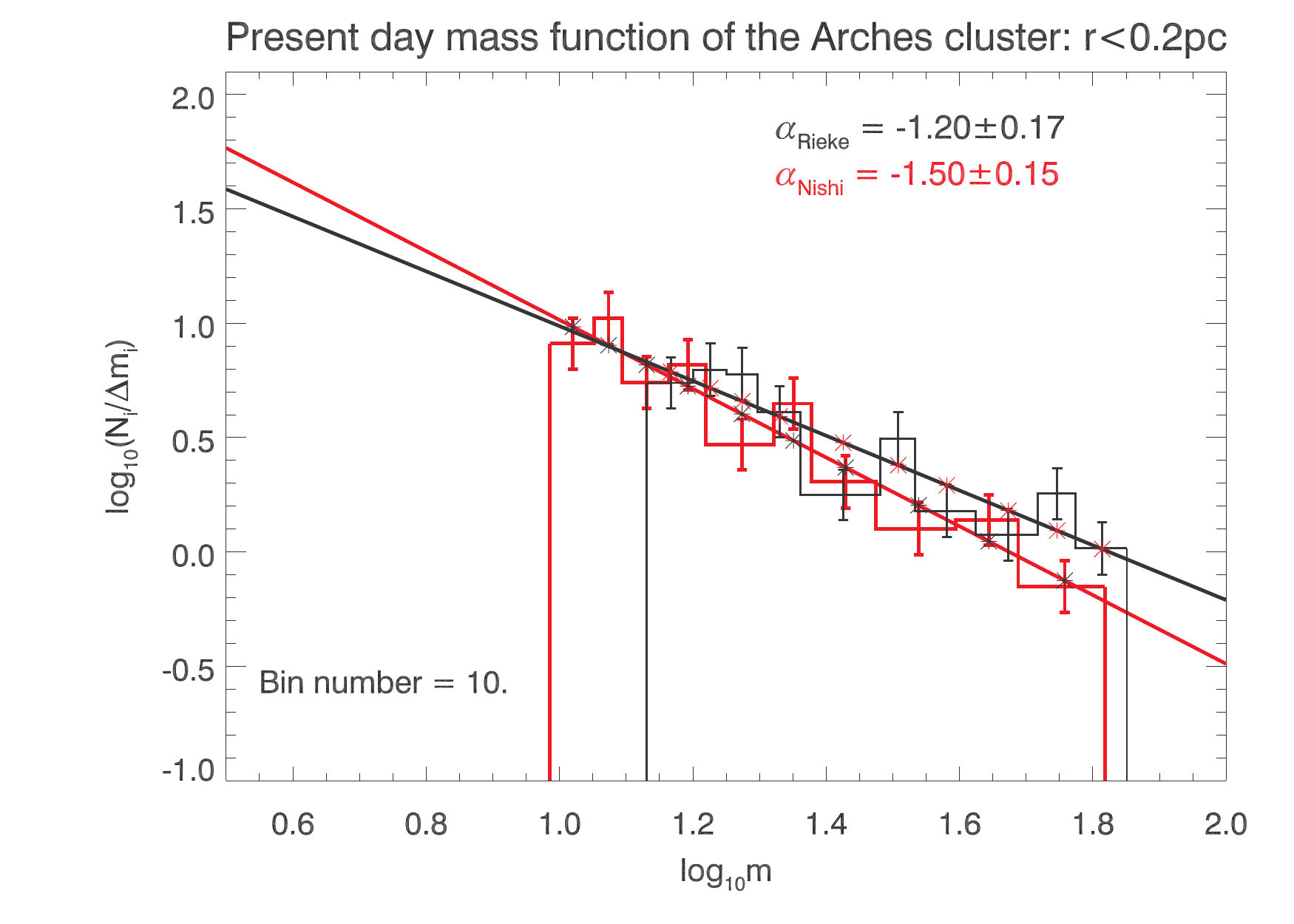} &
\includegraphics[trim=9mm 5mm 0mm 0mm, clip,scale=0.55]{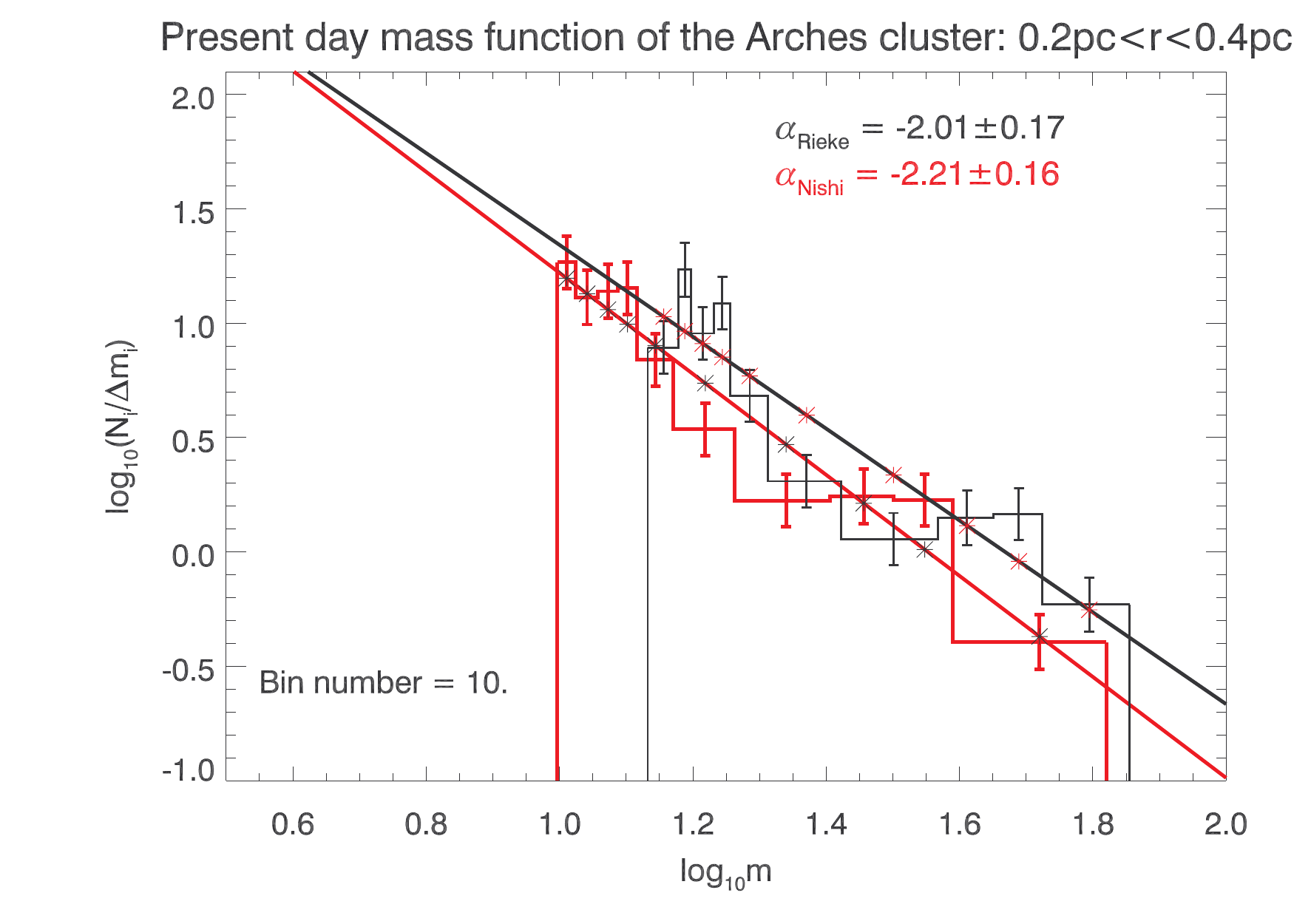} \\
\includegraphics[trim=9mm 5mm 0mm 0mm, clip,scale=0.55]{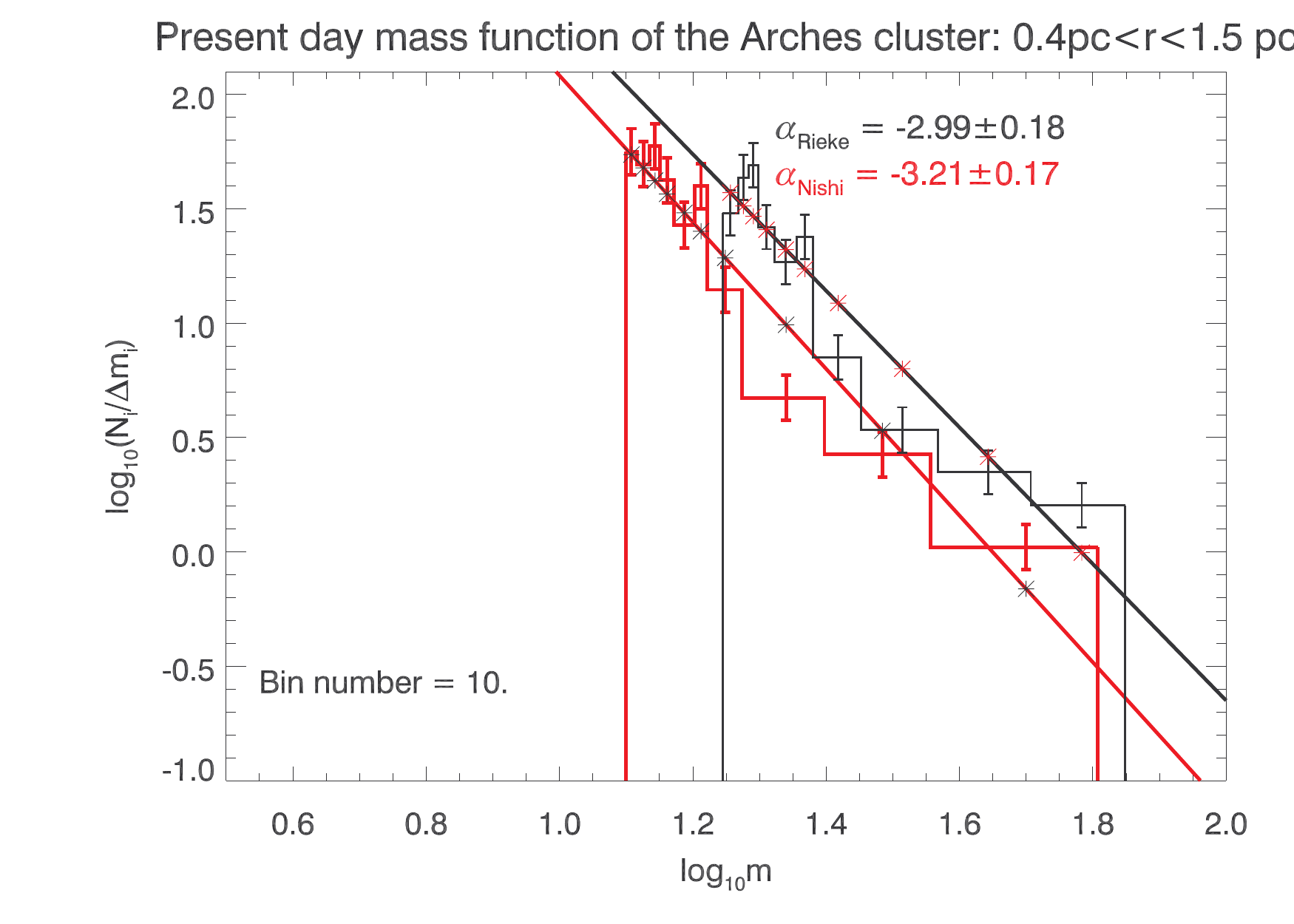} &
\includegraphics[trim=9mm 5mm 0mm 0mm, clip,scale=0.55]{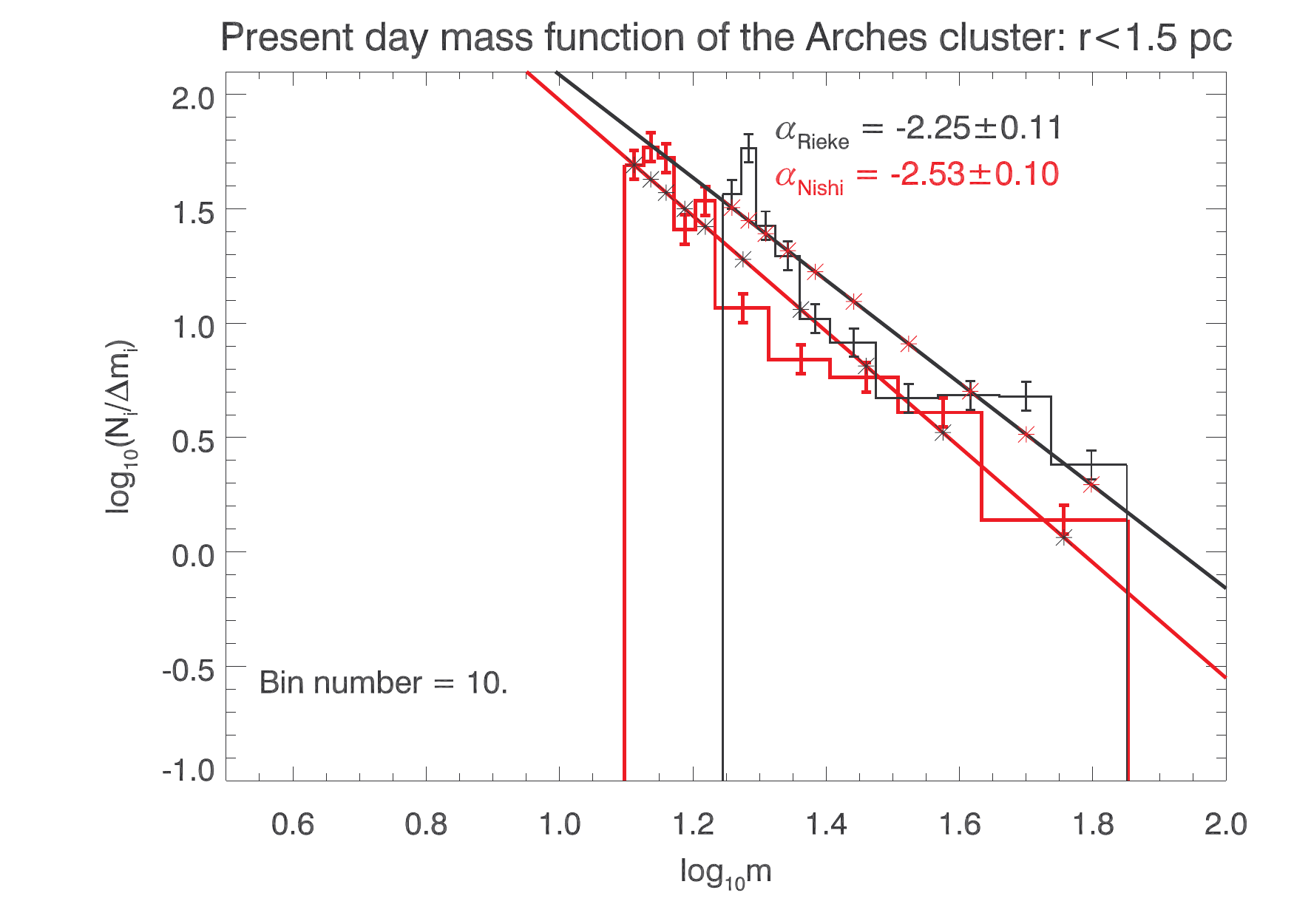} \\
\end{array}$
\caption{The present-day mass function of the Arches cluster. Red lines correspond to the mass distribution derived based on the N-EL, while black lines represent the mass function  assuming the RL-EL. Both mass functions are fitted with a power-law like function with reported slopes of $\alpha$ shown in the respective color. The mass functions are plotted in three regions: (a) the inner core of $r<0.2 $ pc, (b) the intermediate annulus of $0.2 <r<0.4 $ pc and, (c) the cluster outskirts of $0.4 <r<1.5 $ pc. The mass function steepens as we move outward from the cluster center. The complete mass distribution of the cluster within $r<1.5 $ pc (d) is consistent within the uncertainties with a Salpeter IMF. The illustrated error for the slopes only represents the numerical fitting uncertainties (see Table \ref{slopes}, and Sect.~5.2 for a discussion).}

\label{MF}
\end{figure*}

To avoid the bias resulting from the number of stars per bin and its assigned weight in the deriviation of the PDMF slope, we utilize the binning method described by Ma\' iz Apell\' aniz \& Ubeda \cite{maiz}. Following this method, we allow for dynamic bin sizes such that each bin contains approximately the same number of stars, hence the same statistical weight (for details about the implementation of this method see Hu{\ss}mann et al. \citealt{benjamin}).

Figure \ref{MF} shows the derived PDMFs of the Arches cluster on logarithmic scale such that a standard Salpeter mass function \cite{Salpeter} is a line with a slope of $\alpha=-2.35$. The PDMFs are represented well by a single slope power-law function ($\frac{dN}{dm}\propto m^{-\alpha}$) at all radii. The PDMF of the cluster as plotted in three different regions: the inner core with $r<0.2  $ pc, the intermediate annulus of $0.2 <r<0.4$ pc, and the outskirts  which cover partly the outer annulus with $0.4<r<1.5$ pc. Since our fields (see Fig. \ref{fields}) do not cover the whole area of the outer annulus, $0.4<r<1.5$ pc,  the number of sources in each mass-bin is scaled to the observed ratio of the outer annulus area. The projected distances of 0.2 pc and 0.4 pc from the cluster center are the estimated core and half mass radius of the cluster, respectively (Figer, McLean \& Morris \citealt{Figer-McLean1999}; Stolte et al.\citealt{stolte_2002}; Harfst et al. \citealt{harfst2010}). Most of the previous studies have used these annuli to report the mass function slopes. The radius of $1.5$ pc is chosen to cover the cluster out to its tidal radius. For a detailed discussion about the tidal radius of the Arches cluster see Sect. \ref{tidal-radius}.

The derived slopes using the two extinction laws (Fig. \ref{MF}) show that changes in the slope of the fitted power-law function due to the extinction law are similar to the fitting uncertainty (for the detailed description of the systematic and the random errors of the derived slopes see Sect. \ref{error-PDMF}). Kim et al. \cite{kim2006} report almost no change in the slope of the mass function using the two extinction laws of Nishiyama et al. \cite{Nishi2006} and of Rieke, Rieke, \& Paul \cite{rrp} for a mass range that starts at lower masses, $1.3 M_{\odot}<M<50  M_{\odot}$. Since they convert $K_{s}$ magnitudes into stellar masses using an average extinction value for the ensemble of stars with the justified argument that their radial coverage is limited to $\sim 0.2-0.35$ pc (compare to our coverage of up to 1.5 pc), the effect from the extinction law slope is not expected to be large.

 The choice of the extinction law applied for the individual dereddening of each star alters the shape, particularly the lower and the upper mass limits of the resulting MF.  Changing the boundaries of the mass function translates to shifting the highest observed mass to lower values. This is crucial because the Arches cluster is not expected to have any supernova at its present age. Since the cluster is believed to cover the full mass range, it was used to derive a possible upper mass limit of $M=150 M_{\bigodot}$ for the star formation process in the Milky Way (Figer \citealt{Figer2005}). Such an upper mass limit has severe implications for our understanding of the stellar evolution and the formation of the highest mass stars. While the most massive initial mass in our sample is 104$M_{\bigodot}$ when dereddening with a RL-EL, the highest initial mass is only 80$M_{\bigodot}$ when the N-EL is used. This suggests that the claimed upper mass strongly depends on the choice of the extinction law and should be revisited with the steeper extinction laws found toward the GC in the past few years. Such an investigation is beyond the scope of this paper.

 We obtain a PDMF slope of $\alpha_{Nishi}=-1.50\pm0.35$ in the cluster core ($r<0.2$ pc) (see Fig. \ref{MF}, upper left). In the  intermediate annulus with $0.2<r<0.4$ pc, the slope of the PDMF reaches $\alpha_{Nishi}=-2.21 \pm0.27$, which is in the range of the standard Salpeter mass function \cite{Salpeter} with $\alpha=-2.35$, (Fig. \ref{MF}, upper right). The number of massive stars compared to low-mass stars continues to decrease on the outskirts of the cluster. The PDMF for the outer annulus of the cluster, $0.4<r<1.5$ pc, is depleted of high-mass stars with a slope of $\alpha_{Nishi}=-3.21 \pm0.30$ (Fig. \ref{MF}, lower left).
 
   A flattening of the Arches mass function toward the center has been shown in previous studies.  Stolte et al.  \cite{stolte_2005} find a slope of -1.26 for the mass range of $10M_{\bigodot}<M<63 M_{\bigodot}$ in the inner core.  This study corrected for a systematic radial extinction variation. Individual extinction corrections by Espinoza et al. \cite{espinoza} resulted in a slope of -1.88  in the cluster core 
   \footnote{Espinoza et al.  \cite{espinoza} used initial instead of present-day stellar masses to derive the mass function.  The comparable slopes to Espinoza et al.  \cite{espinoza} (derived from initial masses) are shown for comparison in Table \ref{slopes}.}.
 
Our finding, $\alpha_{Nishi}=-2.21 \pm0.27$, in the intermediate annulus is consistent with the slope of $\alpha = -2.28$ found by Espinoza et al. \cite{espinoza} in the same region for stars above $10 M_{\bigodot}$. Kim et al. \cite{kim2006} corrected for a single extinction value and found a flatter slope of -1.71 in this annulus ($0.2<r<0.35$ pc) only for the stars in their high-mass range  ($5 M_{\bigodot}<M<50M_{\bigodot}$). 
 
\begin{figure*}
   \centering
  \includegraphics[scale=0.8]{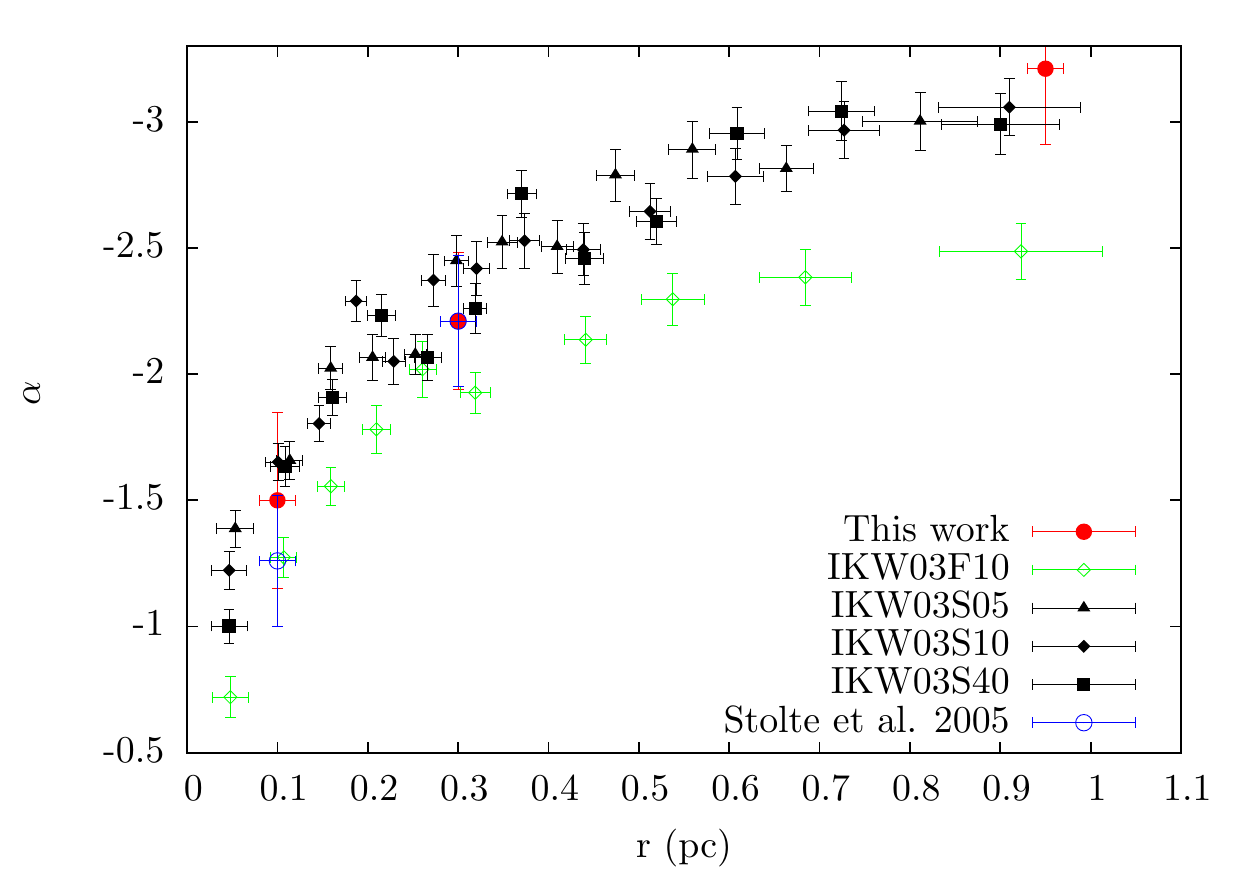}
   \caption{The figure is adopted from Harfst et al. \cite{harfst2010} in their Fig. 13, and compares the mass function slopes from the best-fitting models of N-body simulations of the Arches cluster to the observed values of this work and also from Stolte et al. \cite{stolte_2005}. The black filled symbols represent the models with a Salpeter IMF with different lower mass limits, while green open symbols correspond to a model with a flat IMF.  The models  deviate primarily at larger radii ($r>0.4$ pc). The derived slope from a model starting with a Salpeter IMF at birth in the radius of 1 pc is $\alpha \sim -3$, which is in good agreement with our finding of $\alpha_{Nishi} \sim -3.21 \pm 0.30$ in the outskirts of the Arches cluster. }
\label{harfst-plot}
\end{figure*}

Since is no reported slope for the outskirts of the cluster in these previous studies, we compare our result with simulations. Harfst et al. \cite{harfst2010} performed a series of N-body simulations to find the best model and initial conditions to reproduce the observed data of the Arches cluster. They constructed the mass function for the best fitting models, which have a King model concentration parameter of $W_{0} = 3$. They consist of one model with a flat initial mass function (IMF) and three with a Salpeter IMF (see Fig. \ref{harfst-plot}). While the models with the Salpeter IMF are more consistent with the observed slopes, all the models exhibit a flattening toward the center. The three Salpeter and the one flat IMF model  deviate primarily at larger radii ($r>0.4$ pc) where the predicted slope difference is $\sim 0.5$ dex. The derived slope from a Salpeter IMF model by Harfst et al \cite{harfst2010} at the radius of 1 pc is $\alpha \sim -3$, which is in very good agreement with our finding of $\alpha_{Nishi} \sim -3.21 \pm 0.30$ on the outskirts of the cluster (Fig. \ref{harfst-plot}). This picture is consistent with the dynamical evolution of the Arches cluster as the origin of its observed mass segregation,
which implies that primordial mass segregation is not required to explain the spatial variation in the mass function slope of the Arches cluster. 
  Obtaining the combined PDMF including all the sources within our fields up to $\sim 1.5  $ pc yields a slope of $\alpha_{Nishi}=-2.53 \pm0.31$, similar to the Salpeter IMF within the uncertainties (see Fig. \ref{MF}).

Integrating the PDMF across the desired mass range yields the total mass of the cluster. The complete PDMF ($r<1.5$ pc) of the cluster with a slope of $\alpha_{Nishi}= -2.53 \pm 0.31$ was integrated over the mass range of 1 - 66 $M_{\bigodot}$,  yielding a total mass of $M_{cl}=(1.9^{+0.3} _{-0.3} ) \times 10^4 M_{\bigodot}$ for the Arches cluster. This value is comparable within the uncertainty to the dynamical mass estimate of the cluster. Clarkson  et al. \cite{Clarkson} measured the dynamical mass from the cluster's velocity dispersion to be $M_{cl} (r < 1.0 pc) = 1.5^{+0.74} _{-0.6} \times 10^{4} M_{\bigodot} $ for the Arches cluster. The dynamical mass of the cluster is derived from the velocity dispersion of high- to intermediate-mass stars in the cluster core ($r < 0.2$ pc), which might underestimate the total cluster mass due to the lower velocity dispersion as a consequence of mass segregation.
Previous studies estimated  about two times higher photometric masses for the cluster. However, their results are not directly comparable to our numbers since they were calculated over a narrower annulus (eg. Espinoza et al. \citealt{espinoza}). 
The discrepancy can be due to an extrapolation that is based on the slopes derived from the center of the cluster ($r< 0.4$ pc) and might have  resulted in overestimating the number of higher mass sources in the outskirts. 

\begin{table*}
\caption{Source list. From left to right the columns are: Sequential ID for stars, R.A. offset from the reference star which is the source with ID 1 (R.A.=17:45:50.046 , Dec. = -28:49:23.62), DEC. offset from the reference star, measured $J$-band, $H$-band, and $K_{s}$-band brightness, along with the photometric uncertainty in $K_s$ (columns 4-7), estimated $K_{s}$-band extinction by applying the RL-EL (column 8) and the N-EL (column 9), present-day mass  applying the RL-EL (column 10) and the N-EL (column 11), initial mass applying the N-EL (column 12), and the cluster field in which the detected source is located (column 13). When one of the above values is not available for a source it is denoted as -9999 in the table.  }             
\label{source_list}      
\centering          
\begin{tabular}{c c c c c c c  c c c c c c c} 
\hline\hline       
     
ID & $\triangle R.A.$ & $\triangle Dec.$  & J & H  &$K_{s}$ &$\sigma_{K_{s}}$&$A_{K_{s}(RL-EL)}$ & $A_{K_{s}(N-EL)}$ &  $PD-M_{RL-EL}$& $PD-M_{N-EL}$ & $I-M_{N-EL}$& field\\ 
 & (sec)  & (sec)  & (mag)& (mag) & (mag) && (mag)     & (mag) & $M_{\bigodot}$ & $M_{\bigodot}$ &$M_{\bigodot}$& \\ 

\hline                    
1 &   0     & 0    & 16.79 & 13.59 & 11.54 & 0.007& 3.47 & 2.67 & 61 & 52 & 54 &   1\\
2 & 10.261&-2.708 & -9999 & 12.38 & 10.25 & 0.005 & 3.58 & 2.76 & 71 & 65 & 78 &   1\\
3 & 6.817 & 6.062 & 14.89 & 12.16 & 10.25 & 0.008 & 3.68 & 2.84 & 71 & 65 & 77 &   1\\
4 & 2.071 & 1.391 & 15.63 & 12.37 & 10.29 & 0.013 & 3.55 & 2.74 & 68 & 62 & 71 &   1\\
5 & 4.487 & 2.395 & 15.39 & 12.41 & 10.43 & 0.007 & 3.34 & 2.58 & 65 & 61 & 67 &   1\\

\hline                  
\end{tabular}

\begin{list}{}{}
\item[] The complete table of the sources will be published in the electronic edition of the A\& A journal.
\end{list}
\end{table*}

\subsection{Tidal radius of the cluster}\label{tidal-radius}
The projected distance of the Arches cluster is $\sim 26$ pc from the GC. However, since the Galactic mass distribution and the  location of the cluster along the line of sight has not been determined, it is not easy to estimate the true tidal radius of the cluster. Kim et al. \cite{kim2000} estimated the tidal radius of the cluster based on  a power-law approximation of Galactic mass inside a Galactocentric radius, $r_{g}$.  Their comparison between the simulated  projected number density and the observed mass functions at the time did not rule out a Galactic distance, $r_{g}$,  of 20 and 50 pc for the Arches cluster, but favored  $r_{g}=30$ pc, which results in a tidal radius of $\sim 1$ pc. A similar study by Portegies Zwart et al. \cite{Portegies} concludes that to reproduce the observed density within the half-mass radius, the Arches cluster should lie at the distance of  50-90 pc from the Galactic center, which results in a tidal radius of 1.6-2.5 pc (see Portegies Zwart et al. \cite{Portegies}, their Table 4).

A more recent study by Launhardt et al. \cite{Launhardt}  estimates the enclosed mass in the inner 500 pc of the Milky way. This observational study  shows that the potential in the inner Galaxy cannot be modeled with a single power law and takes  the mass contribution  of the nuclear stellar cluster, the nuclear stellar disk, the central black hole, the Galactic bulge, and the interstellar mass in the nuclear bulge into account(see Launhardt et al. \citealt{Launhardt}, Fig. 14).  Applying the enclosed mass from the Launhardt et al. \cite{Launhardt} model, $M_{g}$, for the Galactocentric range of  $r_{g}=$30-100 pc and a cluster mass of $M_{cl} \sim 19000 M_{\bigodot}$ yields a tidal radius, $r_{t}$, of $1.6-2.5$ pc:

$r_{t}=(\frac{M_{cl}}{2 \times M_{g}})^{1/3}\times {r_{g}}\simeq 1.6-2.5 $ pc.

We observed the Arches cluster out to 1.5 pc, which is close to the range of the derived tidal radii.

\subsection{Uncertainties of the mass function slopes}\label{error-PDMF}
The uncertainties of the PDMF slopes  illustrated in Fig. \ref{MF} represent the numerical fitting uncertainties ($\sigma_{fit} \sim 0.15$). There are also systematic errors that contribute to the final estimated Uncertainties. One source of systematic uncertainty originates from the cluster age, metallicity, and the corresponding choice of the isochrone.
Studies determining the age of the Arches cluster are based on spectroscopy of nitrogen-rich (WN) Wolf-Rayet stars (WRs) and O super-giants, along with the absence of carbon-rich WR stars that establish an upper limit for the age of the cluster. Although studies agree on a narrow evolutionary spread and subsequently a narrow age spread among the cluster stars, it is still hard to confine the absolute age of the Arches cluster. Blum et al. \cite{blum} studied 2 $\mu$m narrow-band images and derived an age of 2-4 Myr assuming the stars are of WR type. Najarro et al. \cite{Najarro} analysed spectra of five massive stars (classified as  3 WNLs and 2 O stars) and obtained an age of 2-2.5 Myr. They derived  solar metallicity for all the five sources based on measuring their N abundances.  More recent work by Martins et al.  \cite{martins} used $K$-band spectra of 28 bright stars in the cluster. An age of 2-3 Myr was obtained for WN7-9 stars while some O super-giants could have ages up to 4 Myr. They suggested that lighter elements are probably slightly super-solar though iron peak elements show solar metallicity.
Espinoza et al. \cite{espinoza} built  mass functions using different ages (2-3.2 Myr) and show that the uncertainty from the choice of age or metallicity will add an uncertainty of $\sigma_{iso} \sim 0.1$ dex. This result agrees with Hu{\ss}mann et al. \cite{benjamin}, who applied different isochrones (3-5 Myr) to construct the PDMF of the Quintuplet cluster.

The second significant source of uncertainty stems from the choice of the extinction law (see Sect. \ref{sec:exti}). We conclude that if none of the extinction laws can be disregarded based on the scientific discussion (see Sect. \ref{sec:comp} and Fig. \ref{CCD_f1} ), an uncertainty of 0.17 dex should be considered as a result of the choice of the extinction law (see Fig. \ref{MF} and Table \ref{slopes}). Applying the steeper extinction laws (Nishiyama et al. \citealt{Nishi2009}; Schoedel  et al. \citealt{schoedel}; Stead \& Hoare \citealt{Stead}) steepens the slope of the PDMF by $0.03-0.30$ dex (see Fig \ref{MF}). It is important to notice that this steepening is a systematic uncertainty. Although all the sources of random and systematic errors are detached and reported above, for the sake of comparison with previous studies with different assumptions about the  extinction law and the age of the cluster, we approximate the mean overall uncertainty in the PDMF slope as a quadratic sum of random and systematic errors of $\sim 0.24$ dex as a result of all the uncertainties discussed above ($\sigma_{fit}, \sigma_{iso}, \sigma_{EL}$). The individual overall uncertainties  are shown in Table \ref{slopes}.

\section{Conclusions}\label{sec:conc}

 We derive the present-day mass function of the outskirts of the Arches cluster for the first time to obtain a full understanding of the cluster mass distribution up to its tidal radius. The $K_{s}$ and $H$-band AO images taken with the VLT as well as $J$-band images obtained with Subaru are analyzed for this investigation. We also present a quantitative study on the effect of the applied extinction law  on the derivation of the mass function in young massive star clusters using the Arches cluster as a template. We show how the choice of the near-infrared extinction law influences the absolute values of the patchy extinction toward the star clusters in the Galactic center, and how it affects the slope of the mass function. The main findings of our study are

 \begin{enumerate}
  \item The extinction law derived by Rieke \& Lebofsky \cite{rieke1985}, $A_{\lambda} \propto \lambda^{-1.61}$, was used commonly in previous studies toward the Arches cluster. We compare the results obtained from Rieke \& Lebofsky \cite{rieke1985} to the results acquired assuming the extinction law derived by  Nishiyama et al. \cite{Nishi2009}, as several recent studies toward the GC suggest a steeper wavelength dependence of $A_{\lambda} \propto \lambda^{-2}$  for the near-infrared extinction law. Applying these different extinction laws results in an average systematic extinction difference of $\Delta A_{K_{s}}=0.8$ mag ($\sim 24\%$), which can reach up to 1.1 mag. The obtained average extinction values applying the Nishiyama et al. EL are $2.5 \pm 0.2$, $2.6\pm 0.2$, and  $2.6\pm 0.3$ mag for the three investigated cluster annuli of $r<0.2\;$ pc, $0.2 <r<0.4\;$ pc, and $0.4 <r<\sim 1.5\;$ pc, respectively. Applying the Rieke \& Lebofsky EL increases the average extinction values to  $3.3 \pm 0.3$, $3.4 \pm 0.3$, and $3.5 \pm 0.4$ mag using the same sample of cluster stars.
      \item The extinction map derived by individual dereddening of the cluster member candidates  shows a high and variable extinction across the cluster.  The derived extinction values applying the Nishiyama et al. \cite{Nishi2009} extinction law  and Rieke \& Lebofsky \cite{rieke1985} vary within the range of $2 < A_{K_{s}} <3.4$ mag and $2.7 < A_{K_{s}} <4.5$ mag, respectively.  The patchiness of the extinction is high and hardly follows any trend; nevertheless, a region of relatively low extinction is present toward the center of the cluster. An area of high extinction is located in the southwest and partly northwest of the cluster. 
     
       \item Obtained stellar masses assuming the Nishiyama et al. \cite{Nishi2009} extinction law are $\sim 30\%$ less massive than derived masses assuming the formerly used extinction law by Rieke \& Lebofsky \cite{rieke1985}. Considering that the Arches cluster hosts a collection of high-mass stars, a mass difference of $\sim30\%$ can change the estimated individual initial masses substantially. In this work, the initial mass of the most massive star in our sample decreases from 104 $M_{\bigodot}$ to 80 $M_{\bigodot}$ when using the steeper extinction law, which has severe implications for discussion of the possible maximum stellar mass forming in the Milky Way (see Sect. \ref{sec:mf} for a detailed discussion).

       \item We derive the present-day mass function of the Arches cluster for an area that  reaches out to the tidal radius. Our mass functions cover the high-mass  part of the mass spectrum (see Table \ref{mass_cut_info}) and are obtained in three different annuli of $r<0.2 $ pc (core), $0.2<r<0.4$ pc (intermediate annulus), and $0.4<r< 1.5$ pc (outer annulus). The slopes derived by assuming the Nishiyama et al. \cite{Nishi2009} extinction law increase from a flat slope of $\alpha_{Nishi}=-1.50 \pm0.35$ in the core to $\alpha_{Nishi}=-2.21 \pm0.27$ in the intermediate annulus, which steepens to $\alpha_{Nishi}=-3.21 \pm0.30$ in the outer annulus. It is important to notice that uncertainties from the assumption of the extinction law and the age of the cluster are systematic. The slope derived by applying the Rieke \& Lebofsky \cite{rieke1985} EL are on average flatter by 0.17 dex. Our mass functions show that, while in the core of the cluster the number of massive stars is greater than lower mass ones when comparing to the normal Salpeter mass function ($\alpha=-2.35$, \citealt{Salpeter}), the intermediate annulus has a Salpeter-like distribution, and the outskirts are depleted of high-mass stars. Comparing the mass function slopes  of the cluster with the N-body simulations performed by Harfst et al. \cite{harfst2010}, we conclude that the radial variation observed in the present-day mass function slopes is consistent with dynamical evolution  of a cluster that formed with a normal Salpeter IMF.           
       
 \item
 The extrapolation of the complete PDMF for the area of $r<1.5$ pc, derived for a present-day mass range of 1-66 $M_{\bigodot}$  and a slope of $\alpha_{Nishi}=-2.53 \pm0.31$, results in a total mass of $M_{cl}=(1.9^{+0.3} _{-0.3} ) \times 10^4 M_{\bigodot}$ for the Arches cluster.
   
   \end{enumerate}

It was argued before that  the IMF might be top-heavy in the GC because of increased cloud temperatures and magnetic fields (Morris \citealt{morris}, Dib et al. \citealt{sami_dib}, see also Klessen et al. \citealt{klessen} and Stolte et al. \citealt{stolte_2005} for discussions). The slopes derived in our analysis are consistent with the expectations of the dynamical evolution of the cluster starting with a Salpeter IMF at birth. We therefore do not need to invoke a primordially deviating mass function  to explain the spatial distribution of stars in the Arches cluster.

\begin{acknowledgements}
       We thank the anonymous referee for providing careful comments that helped improve the paper. M.H., B.H., and A.S. acknowledge funding from the German science foundation (DFG) Emmy Noether program under grant STO 496-3/1, and we thank the Argelander-Institute for Astronomy at the University of Bonn for providing such a scientifically stimulating atmosphere. We also thank MPIA for generous support during continuous visits and the fruitful scientific exchange. It is a pleasure to thank the NAOS-CONICA instrument team for their extensive support during the observations.
\end{acknowledgements}

\end{document}